\documentclass[12pt,aps,showpacs]{revtex4}
\usepackage{epsfig}
\usepackage{bbm}
\usepackage{amssymb}
\usepackage{amsmath}

\newcommand{\lbfig}[1]{\refstepcounter{fig} \label{#1} }
\newcounter{fig}
\newcommand{\nc}{\newcommand}
\nc{\be}{\begin{equation}}
\nc{\ee}{\end{equation}}
\nc{\bea}{\begin{eqnarray}}
\nc{\eea}{\end{eqnarray}}
\nc{\nn}{\nonumber}
\nc{\acom}[2]{ \left\{ #1,#2 \right\} }
\nc{\com}[2]{ \left[ #1,#2 \right] }
\nc{\dd}{^\dagger}
\nc{\ddp}{^{\dagger\prime}}
\nc{\ddpp}{^{\dagger\prime\prime}}
\nc{\pp}{^{\prime\prime}}
\nc{\ml}{M^\dagger M}
\nc{\mr}{MM\dd}
\nc{\explrp}{{\rm e}^{\frac{i}{2} \overset{\leftharpoonup}{\partial_z}
  \overset{\rightharpoonup}{\partial_{k_z}} }}
\nc{\exprlp}{{\rm e}^{\frac{i}{2} \overset{\leftharpoonup}{\partial_{k_z}} 
  \overset{\rightharpoonup}{\partial_z} }}
\nc{\explrm}{ e^{-\frac{i}{2} \overset{\leftharpoonup}{\partial_z}
  \overset{\rightharpoonup}{\partial_{k_z}} }}
\nc{\exprlm}{ e^{-\frac{i}{2} \overset{\leftharpoonup}{\partial_{k_z}} 
  \overset{\rightharpoonup}{\partial_z} }}
\nc{\lp}{\left(}
\nc{\rp}{\right)}
\nc{\CP}{{\cal CP}}
\nc{\CPd}{({\cal CP})^\dagger}
\nc{\Q}{{\cal Q}}
\nc{\Qd}{({\cal Q})^\dagger}

\def\Slash#1{#1\kern-0.55em\raise.05ex\hbox{/}}
\def\slash#1{#1\kern-0.5em\raise.05ex\hbox{{$\scriptstyle /$}}}
\newcommand{\deldag}{\mathbin{\partial\mkern-10.5mu\big/}}

\begin{document}
\rightline{HD-THEP-04-36, ITP-UU-04/21, SPIN-UU-04/12}


\title{Kinetic description of fermion flavor mixing and 
       CP-violating sources for baryogenesis}


\author{Thomas Konstandin$^*$, Tomislav Prokopec$^{(1)*}$
                       and 
                   Michael G. Schmidt
       }
\email[]{T.Konstandin@thphys.uni-heidelberg.de}
\email[]{T.Prokopec@phys.uu.nl}
 
\email[]{M.G.Schmidt@thphys.uni-heidelberg.de}

\affiliation{Institut f\"ur Theoretische Physik, Heidelberg University,
             Philosophenweg 16, D-69120 Heidelberg, Germany}

\affiliation{$(1)\,$ 
             Institute for Theoretical Physics (ITF) \& Spinoza Institute,
             Utrecht University, Leuvenlaan 4, Postbus 80.195,
             3508 TD Utrecht, The Netherlands}

\date{\today}
\begin{abstract}
We derive transport equations for fermionic systems with a space-time
dependent mass matrix in flavor space allowing for complex elements
leading to CP violation required for electroweak baryogenesis.
By constructing appropriate projectors in flavor space of the
relevant tree level Kadanoff-Baym equations, we split the 
constraint equations into "diagonal" and "transversal" parts
in flavor space, and show that they decouple. While the diagonal
densities exhibit standard dispersion relations at leading order
in gradients, the transverse densities exhibit a novel on-shell structure. 
Next, the kinetic equations are considered to second order in gradients and 
the CP-violating source terms are isolated. This requires a thorough discussion
of a flavor independent definition of charge-parity symmetry operation.
To make a link with baryogenesis in the supersymmetric extension
of the Standard Model, we construct the Green functions for the leading 
order kinetic operator and solve the kinetic equations for two mixing fermions
(charginos). We take account of flavor blind damping, 
and consider the cases of inefficient and moderate diffusion. 
The resulting densities are the CP-violating chargino currents
that source baryogenesis.  
 
\end{abstract}

\pacs{
98.80.Cq,  
11.30.Er,  
11.30.Fs   
} 

\maketitle

%
%

\section{Introduction}

The problem of fermion mixing in kinetic theory is important both
for electroweak scale baryogenesis, as well as for neutrinos. 
In the case of baryogenesis one deals with the dynamics of chiral fermions
at the electroweak phase transition, with a mass matrix generated by 
a space-time dependent Higgs condensate. For studies of baryogenesis 
in the Minimal Standard Model the relevant mixing occurs between
the standard model
quarks~\cite{KonstandinProkopecSchmidt:2003,Farrar:sp}.
For baryogenesis in two Higgs doublet models the relevant mixing is in 
the Higgs sector~\cite{JoyceProkopecTurok:1995,JoyceProkopecTurok:1994-I,
JoyceProkopecTurok:1996}. In supersymmetric extensions of the Standard Model
the mixing is in the chargino and neutralino sectors
\cite{HuetNelson:1996,DaviesFroggattMoorhouse:1996,
CarenaQuirosRiottoViljaWagner:1997,ClineJoyceKainulainen:1998,
CarenaMorenoQuirosSecoWagner:2000,ClineJoyceKainulainen:2000+2001,
KainulainenProkopecSchmidtWeinstock:2001,
KainulainenProkopecSchmidtWeinstock:2002,
HuberSchmidt:2000+2001,CarenaQuirosSecoWagner:2002,
ProkopecSchmidtWeinstock:2003,ProkopecSchmidtWeinstock:2004}.

We consider the dynamics of chiral fermions interacting with a plasma
according to the Lagrangian, 
\begin{eqnarray}
  {\cal L}_f &=&
              i \bar{\psi} \deldag \psi
            - \bar{\psi}_L m \psi_R - \bar{\psi}_R m^* \psi_L
            + {\cal L}_{\rm int}
\,.
\label{lagrangian}
\end{eqnarray}
where $m=m_h + i m_a$ is the fermion mass matrix and 
${\cal L}_{\rm int}$ specifies interactions of fermions with the rest of 
the plasma. Since we are interested in a near equilibrium
dynamics of fermions at the phase transition where they acquire the mass
through a Higgs mechanism, we shall assume that the mass is space-time 
dependent. Moreover, we shall assume that a `thick wall' approximation 
applies, in the sense that the typical momenta of particles are much larger 
than the rate of change of the background, $\partial^\mu m \ll k^\mu m$.  
Since we are mostly interested in electroweak scale baryogenesis, 
we shall assume that ${\cal L}_f$ violates CP symmetry 
either through complex elements of $m_{h,a}$, or through interactions
${\cal L}_{\rm int}$ (complex Yukawas, {\it etc.}).

Due to space-time dependence of 
the mass matrix, CP violation, which is crucial for baryogenesis,
is present already in the case of two fermion mixing. 

Basis of our discussion as in
Refs.~\cite{ProkopecSchmidtWeinstock:2003,ProkopecSchmidtWeinstock:2004} 
are the Kadanoff-Baym equations for fermionic Wightman functions.
They are Wigner transformed such that one can eventually detect 
semiclassical features in the resulting transport equations. 
In the case of several mixing flavors, semiclassical
(quasiparticle) dynamics does not necessarily lead to an
accurate description of the kinetics. In this work we
investigate the kinetics of mixing fermions in gradient approximation, 
but without resorting to a semiclassical approximation. 

In Refs.~\cite{ProkopecSchmidtWeinstock:2003,ProkopecSchmidtWeinstock:2004}
such quasiclassical behavior was found at order $\hbar$ in gradient expansion.
The semiclassical behavior was argued on the basis of a derivative expansion 
and a separation of the basic equations into constraint and transport 
equations. The latter turned out to be
naturally consistent and allowed reduction to transport equations 
for the spin dependent on shell distribution functions in the 
position dependent mass eigenbasis. Exept in the case of a
near mass degeneracy, the dynamics of nondiagonal elements 
of the Wightman function could be shown not to influence the 
dynamics of the physically interesting diagonal ones to order $\hbar$ 
of the derivative expansion. In this work we relax this 
limitation combining the basic equations differently and we are able 
to discuss the nondiagonal Wightman functions and their highly nontrivial
constraint equations, which in general are not algebraic.
The resulting formalism does not depend on a particular 
choice of basis, resolving thus the principal limitation of the 
formalism presented in
Refs.~\cite{ProkopecSchmidtWeinstock:2003,ProkopecSchmidtWeinstock:2004}.

A similar attempt within the same Schwinger-Keldysh formalism has been made
in~\cite{CarenaMorenoQuirosSecoWagner:2000,CarenaQuirosSecoWagner:2002}.
The principal limitation of that work is that the projection
onto the diagonal densities is made before the relevant charges are allowed
to propagate, making thus transport of mixing fermions unfeasible.
Moreover, the CP-violating currents are inserted as sources
into the transport diffusion equations in a phenomenological manner,
which does not come out of the Schwinger-Keldysh formalism. 

 This work bypasses both of these limitations, albeit in a
somewhat simplistic disguise. Namely, we assume damping to be flavor blind and 
momentum independent, having as a consequence the following two limitations.
First, we cannot model different damping rates of diagonal and
off-diagonal densities, which may be of crucial importance for proper tracking
of flavor decoherence. Second, taking a momentum independent damping
may give a na\"\i ve picture of transport (diffusion) of CP-violating 
currents.
These limitations are of technical rather than fundamental nature however,
and can be overcome by a more fundamental treatment of collisions,
which is the subject of a forthcoming publication.

\section{Transformation to the Chiral System}

 In order to study the near equilibrium dynamics of fermions in the presence of 
flavor mixing, it is convenient to start with  
the Kadanoff-Baym equations for mixing fermions,
which can be derived from the effective action in the Schwinger-Keldysh
formalism~\cite{ProkopecSchmidtWeinstock:2003}.
When written in Wigner space, the Kadanoff-Baym equations
for the Wightman functions
$iS^{<,>}$ become (with flavor and spinor indices suppressed):
\begin{eqnarray}
 \Big( \slash{k}
          + \frac i2 \deldag 
          - m_h
            {\rm e}^{-\frac{i}{2}\stackrel{\leftarrow}{\partial}
                      \cdot\,\partial_k}
          - i\gamma^5m_a
            {\rm e}^{-\frac{i}{2}\stackrel{\leftarrow}{\partial}
                   \cdot\,\partial_k}
  \Big)S^{<,>}
   -   {\rm e}^{-i\diamond}\{\Sigma_h\}\{ S^{<,>}\}
       -   {\rm e}^{-i\diamond}\{\Sigma^{<,>}\}\{S_h\}
   =       {\cal C}_\psi
\,,
\;\;
\label{Wigner-space:fermionic_eom}
\end{eqnarray}
where $S_h = S^t - (S^>+S^<)/2$, 
 $\Sigma_h = \Sigma^t - (\Sigma^>+\Sigma^<)/2$
($S^t$ and $\Sigma^t$ denote the time ordered (chronological, Feynman) 
Green function and the corresponding self-energy) 
and the collision term reads
\begin{eqnarray}
  {\cal C}_\psi &=&  \frac 12 {\rm e}^{-i\diamond}
                     \big( \{\Sigma^>\}\{S^<\} 
                         - \{\Sigma^<\}\{S^>\}
                     \big)
\,,
\label{Cpsi}
\end{eqnarray}
with $\diamond\{a\}\{b\} \equiv (1/2)(\partial_x a) \cdot \partial_k b
                    - (1/2)(\partial_k a) \cdot \partial_x b$.
These equations are formally exact, provided the self-energies 
$\Sigma^{<,>}$, $\Sigma_a = (i/2)(\Sigma^> - \Sigma^<) \equiv \Gamma_\psi$
and $\Sigma_h$ are calculated exactly. 
(Note that $\Sigma_h$ is also given in terms of $\Sigma^{<,>}$ 
through a spectral relation~\cite{ProkopecSchmidtWeinstock:2003}).
Usually one is lead to a reasonable approximation 
scheme for the self-energies. The one often used is based on 
a truncation of self-energies at a certain loop order
(for an example of a one-loop calculation of the self-energies see
Ref.~\cite{ProkopecSchmidtWeinstock:2003}).
Equations~(\ref{Wigner-space:fermionic_eom}--\ref{Cpsi}) fully
specify the dynamics of fermions, since $S_h$ can be 
determined from $S^{<,>}$ through a spectral relation. 
In some situations a more convenient system
of equations may be the equation for $S^<$ 
(or $F = (1/2)(S^<+S^>)$) together with the equation for the spectral
function ${\cal A} \equiv (i/2)(S^>-S^<)$, which is collisionless, 
\begin{eqnarray}
  \Big( \slash{k}
          + \frac i2 \deldag 
          - m_h
            {\rm e}^{-\frac{i}{2}\stackrel{\leftarrow}{\partial}
                      \cdot\,\partial_k}
          - i\gamma^5m_a
            {\rm e}^{-\frac{i}{2}\stackrel{\leftarrow}{\partial}
                   \cdot\,\partial_k}
  \Big){\cal A}
       -   {\rm e}^{-i\diamond}\{\Sigma_h\}\{ {\cal A}\}
       -   {\rm e}^{-i\diamond}\{\Gamma_{\psi}\}\{S_h\}
   =  0
\,,
\label{spectral-function}
\end{eqnarray}
where $m_h = (1/2)(m+m^\dagger)$ and 
$m_a = (1/2i)(m-m^\dagger)$ denote the hermitean and antihermitean 
parts of the mass matrix $m$, respectively.
Now we assume that ${\cal L}_{\rm int}$ is
governed by a set of weak couplings, and that we are interested in 
nearly equilibrium dynamics of the modes whose momenta and energies are 
not very small (when compared with the temperature of the system), 
such that the perturbative approach is justified. In this approach one 
first solves Eqs.~(\ref{Wigner-space:fermionic_eom}--\ref{spectral-function})
for the thermal tree-level propagators (usually in a gradient expansion),
and then uses these propagators to study the near equilibrium dynamics 
by recasting the equations in a linear response 
approximation, with suitably truncated self-energies.
This approach was pioneered in 
Refs.~\cite{ProkopecSchmidtWeinstock:2003,ProkopecSchmidtWeinstock:2004}.
In this work we focus mostly 
on finding an approximate solution to the tree level dynamics of mixing 
fermions in a gradient expansion. We shall not make any assumption
concerning the eigenvalues of the mass matrix, such that 
our approach is valid also when there are nearly 
degenerate mass eigenvalues, which is not the case
with the approach advocated in Ref.~\cite{ProkopecSchmidtWeinstock:2003},
which applies when
$\hbar^2 k\cdot \partial\ll m_i^2 - m_j^2\; (\,\forall\, i\neq j)$,
where $m_i^2$ are the eigenvalues of the mass matrices squared 
$mm^\dagger$ and $m^\dagger m$.

 For simplicity we assume planar symmetry, such 
that in the wall frame $m=m(z)$, where $z$ denotes the direction along 
which the wall propagates. This assumption is justified when 
the bubbles become sufficiently large~\cite{HuetKajantieLeighLiuMcLerran:1992}.
Working in this frame it is not hard to show that 
the tree level Dirac kinetic operator ${\cal D}$ defined by the 
equation~({\it cf.} Eq.~(\ref{Wigner-space:fermionic_eom}))
\begin{eqnarray}
  {\cal D}iS^{<,>}
  \equiv 
  \Big( \slash{k}
          + \frac i2 \slash{\partial}
          - m_h
           {\rm e}^{\frac{i}{2}\stackrel{\leftarrow}{\partial_z}\partial_{k_z}}
          - i\gamma^5m_a
          {\rm e}^{\frac{i}{2}\stackrel{\leftarrow}{\partial_z}\partial_{k_z}}
  \Big)i S^{<,>}
   = 0
\,,
\label{Wigner-space:fermionic_tree}
\end{eqnarray}
commutes with the spin operator, 
\begin{equation}
 [{\cal D},S_z] = 0\,,\qquad
 S_z = \frac{1}{\tilde k_0}\big(\gamma^0k_0 - \gamma^1k_x-\gamma^2k_y\big) 
       \gamma^3\gamma^5
\,,\qquad \tilde k_0 = {\rm sign}(k_0)(k_0^2-k_x^2-k_y^2)^{1/2}
\,,\quad
\label{Sz}
\end{equation}
provided the coordinate dependences of the Wightman functions are of the 
form $iS^{<,>} = iS^{<,>}\big(k,t-(k_x x+k_y y)/k_0,z\big)$.
In the rest frame of particles $S_z$ measures spin in $z$-direction, such that 
$S_z \stackrel{\rm rest\;frame}{-\!\!\!-\!\!\!-\!\!\!\longrightarrow}
                    \gamma^0\gamma^3\gamma^5$. 
Having found a conserved quantity, 
we can write the solution of~(\ref{Wigner-space:fermionic_tree}) 
in a block-diagonal form in spinor space 
(diagonal in spin)~\cite{ProkopecSchmidtWeinstock:2003}, 
\begin{equation}
 iS^{<,>} = \sum_{s=\pm 1} i S_s^{<,>}
\,,\qquad iS_s^{<,>} = -P_s \big[s\gamma^3\gamma^5 g_0^{s<,>}
                               - s\gamma^3 g_3^{s<,>}
                               + \mathbbm{1} g_1^{s<,>}
                               - i\gamma^5 g_2^{s<,>}\big]
\,,
\label{spin-diagonalS}
\end{equation}
where $P_s = \frac{1}{2} (\mathbbm{1} + s  S_z$ )
is the spin projector, $P_s P_{s'} = \delta_{ss'}P_s$ $(s,s'=-1,1)$,
and $g_0^s$, $g_1^s$, $g_2^s$ and $g_3^s$ denote 
vector, scalar, pseudo-scalar, and pseudo-vector densities of spin $s$
on eight-dimensional phase space $\{k,x\}$, respectively. 

Upon multiplying Eq.~(\ref{Wigner-space:fermionic_tree}) by 
$\{P_s \mathbbm{1} , P_s \gamma_0, - P_s is\gamma_3,-P_s \gamma_5 \}$
and tracing over spinor one finds~\cite{ProkopecSchmidtWeinstock:2003}
\bea
\Big(2i\tilde k_0
      - \frac{k_0\partial_t+\vec k_\|\cdot\nabla_\|}{\tilde k_0}\Big) g_0^s 
   -  \lp 2isk_z 
   + s\partial_z \rp g_3^s 
   - 2im_h \explrp g_1^s
   - 2i m_a \explrp g_2^s
  &=& 0
\quad
\label{B0}
\\
\Big(2i\tilde k_0
      - \frac{k_0\partial_t+\vec k_\|\cdot\nabla_\|}{\tilde k_0}\Big) g_1^s
   -  \lp 2sk_z
   - is\partial_z \rp g_2^s
   - 2im_h \explrp g_0^s 
 \;+\; 2m_a \explrp g_3^s 
  &=& 0
\quad
\label{B1}
\\
\Big(2i\tilde k_0
        - \frac{k_0\partial_t+\vec k_\|\cdot\nabla_\|}{\tilde k_0}\Big) g_2^s
    +  \lp 2sk_z
    - is\partial_z \rp g_1^s 
  \;-\; 2m_h \explrp g_3^s
    - 2i m_a \explrp g_0^s 
   &=& 0
\quad
\label{B2}
\\
\Big(2i\tilde k_0
    - \frac{k_0\partial_t+\vec k_\|\cdot\nabla_\|}{\tilde k_0}\Big) g_3^s
    -  \lp 2isk_z
    + s\partial_z \rp g_0^s 
  \;+\; 2m_h \explrp g_2^s
  \;-\; 2 m_a \explrp g_1^s
   &=& 0
\,,
\qquad
\label{B3} 
\eea
where $\vec k_\|\cdot\nabla_\| = k_x\partial_x + k_y\partial_y$, and we have 
dropped the superscripts $<,>$ of $g_a^s$ $(a=0,1,2,3)$.

 This basis is useful in the one flavor case
(as well as in the mixing case with well separated mass eigenvalues),
since at order $\hbar$ in gradient expansion
the vector density $g_0$ obeys an algebraic constraint equation,
from which one obtains the dispersion relation with a spin dependent 
CP-violating shift appearing at order $\hbar$. 
An important implication of this result is 
that, at order $\hbar$, the quasiparticle picture of the plasma 
is preserved~\cite{KainulainenProkopecSchmidtWeinstock:2001,
KainulainenProkopecSchmidtWeinstock:2002}.  
When inserted into the kinetic equation for $g_0^s$, and upon integration over 
the positive and negative frequencies, one arrives at the Boltzmann-like 
kinetic equation for the distribution function for particles and antiparticles,
respectively, which at second order in gradients (first order in $\hbar$)
exhibits a spin-dependent CP-violating force.

In the case of several flavors however, the basis $g_a^s$ leads to 
mixing between different $g_a^s$'s already at the classical (leading order) 
level. Moreover, $g_a^s$'s do not transform in a definite manner 
under flavor rotations~\cite{ProkopecSchmidtWeinstock:2003}.
A more appropriate basis to describe fermion mixing is the chiral basis
\begin{equation}
 g_R^s = g_0^s + g_3^s
\,,\qquad 
 g_L^s = g_0^s - g_3^s
\,,
\label{gLR}
\end{equation}
and the following densities, 
\begin{equation}
  g^s_N = g_1^s + i g_2^s
\,\qquad
{g^s_N}^\dagger = g_1^s - i g_2^s
\,.
\label{gN:definition}
\end{equation}
These densities do transform in a definite way under mass diagonalization
(flavor rotation), 
\bea
  m\rightarrow m_d = UmV^\dagger
\,,\qquad
  m^\dagger\rightarrow m_d^\dagger = Vm^\dagger U^\dagger = m_d 
\,,\qquad \nn \\
  m m\dd \rightarrow m_d^2 = U m m\dd U^\dagger
\,,\qquad
  m^\dagger m \rightarrow m_d^2 = V m^\dagger m V^\dagger
\,,\qquad
\eea
where the unitary transformation matrices $U$ and $V$
are chosen such that $m_d = m_d^\dagger$ are diagonal mass matrices 
with real eigenvalues 
\bea
  g_L^s \rightarrow g_{Ld}^s &=& U g_L^s U^\dagger
,\qquad
  g_R^s \rightarrow g_{Rd}^s = V g_L^s V^\dagger \nn, \\
\quad
  g_N^s \rightarrow g_{Nd}^s &=& U g_N^s V^\dagger
,\quad
  {g_N^s}^\dagger \rightarrow {g_{Nd}^s}^\dagger = V {g_N^s}^\dagger U^\dagger
.
\label{transformations:gLRN}
\eea

 From Eqs.~(\ref{B0}--\ref{B3}) we easily find
\begin{eqnarray}
\Big(2i\tilde k_0
      - \frac{k_0\partial_t+k_\|\cdot\nabla_\|}{\tilde k_0}\Big) g_R^s 
 \,-\, s\lp 2ik_z + \partial_z \rp g_R^s 
   - 2im^\dagger \hat E g_N^s
  &=& 0
\quad
\label{gR}
\\
\Big(2i\tilde k_0
      - \frac{k_0\partial_t+k_\|\cdot\nabla_\|}{\tilde k_0}\Big) g_L^s 
  \,+\, s\lp 2ik_z + \partial_z \rp g_L^s 
   - 2im \hat E {g_N^s}^\dagger
  &=& 0
\quad
\label{gL}
\\
\Big(2i\tilde k_0
      - \frac{k_0\partial_t+k_\|\cdot\nabla_\|}{\tilde k_0}\Big) g_N^s 
  \,+\, s\lp 2ik_z + \partial_z \rp g_N^s 
  \,-\, 2im \hat E g_R^s
   &=& 0
\,
\quad
\label{gN}
\\
\Big(2i\tilde k_0
    - \frac{k_0\partial_t+k_\|\cdot\nabla_\|}{\tilde k_0}\Big) {g_N^s}^\dagger 
   - s\lp 2ik_z + \partial_z \rp {g_N^s}^\dagger
   - 2im^\dagger \hat E g_L^s
  &=& 0
\,,
\quad
\label{gN+}
\end{eqnarray}
where we introduced the following notation
\begin{equation}
  \hat E \equiv {\rm exp}\Big(\frac i2
                 \overset{\leftarrow}{\partial}_z
                 \overset{\rightarrow}{\partial}_{k_z}\Big)
\,,\qquad
  \hat E^\dagger \equiv {\rm exp}\Big(-\frac i2
                 \overset{\leftarrow}{\partial}_{k_z}
                 \overset{\rightarrow}{\partial}_z\Big)
\,.
\label{EandEdagger}
\end{equation}
Definite transformation properties of these equations are apparent.
Indeed, Eqs.~(\ref{gR}--\ref{gN+}) transform just as 
the densities $g_L^s, g_R^s, g_N^s$ and ${g_N^s}^\dagger$ 
in Eq.~(\ref{transformations:gLRN}).

 From the antihermitean parts of Eqs.~(\ref{gR}--\ref{gL}) we get 
the corresponding constraint equations for $g_R^s$ and $g_L^s$, 
while the constraint equation for ${g_N^s}$ is obtained simply
by taking a hermitean conjugate
of~(\ref{gN+}) and subtracting the result from~(\ref{gN}),
\begin{eqnarray}
   (2\tilde k_0 - 2sk_z) g_R^s 
     - m^\dagger \hat E g_N^s 
     - {g_N^s}^\dagger \hat E^\dagger m
     &=& 0
\label{CE:gR}
\\
   (2 \tilde k_0 + 2sk_z) g_L^s 
     - m \hat E {g_N^s}^\dagger
     - g_N^s \hat E^\dagger m^\dagger
     &=& 0
\label{CE:gL}
\\
   (2 \tilde k_0 - is\partial_z) g_N^s 
    \,-\, m \hat E g_R^s
    \;-\; g_L^s \hat E^\dagger m
    \;&=& 0
\label{CE:gN}
\\
   (2 \tilde k_0 + is\partial_z) {g_N^s}^\dagger
     - m^\dagger\hat E g_L^s
     - g_R^s \hat E^\dagger m^\dagger
     &=& 0
\,.
\label{CE:gN+}
\end{eqnarray}
Note that the constraint equation for
${g_N^s}^\dagger$ is simply a hermitean conjugate of~(\ref{CE:gN}).

 Analogously, the kinetic equations for $g_R^s$ and $g_L^s$
are obtained from the hermitean parts of Eqs.~(\ref{gR}--\ref{gL}), 
while the kinetic equation for ${g_N^s}$ is obtained by 
adding a hermitean conjugate of~(\ref{gN+}) to~(\ref{gN}),
\begin{eqnarray}
- \frac{k_0\partial_t+\vec k_\|\cdot\nabla_\|}{\tilde k_0} g_R^s 
   \;-\; s \partial_z g_R^s 
   - im^\dagger \hat E g_N^s
   + i{g_N^s}^\dagger \hat E^\dagger m
  &=& 0
\quad
\label{KE:gR}
\\
      - \frac{k_0\partial_t+\vec k_\|\cdot\nabla_\|}{\tilde k_0} g_L^s 
 \;+\; s\partial_z g_L^s 
  \; - im \hat E {g_N^s}^\dagger
   + ig_N^s \hat E^\dagger m^\dagger
  &=& 0
\quad
\label{KE:gL}
\\
      - \frac{k_0\partial_t+\vec k_\|\cdot\nabla_\|}{\tilde k_0} g_N^s 
  \;+\, 2isk_z g_N^s 
  \;-\, im \hat E g_R^s
  \,+\, ig_L^s \hat E^\dagger m
   &=& 0
\,
\quad
\label{KE:gN}
\\
  -\frac{k_0\partial_t+\vec k_\|\cdot\nabla_\|}{\tilde k_0} {g_N^s}^\dagger 
   - 2isk_z {g_N^s}^\dagger
   - im^\dagger \hat E g_L^s
   + ig_R^s \hat E^\dagger m^\dagger
  &=& 0
\,
\quad
\label{KE:gN+}
\end{eqnarray}
As above, the equation for ${g_N^s}^\dagger$ is a hermitean conjugate 
of~(\ref{KE:gN}). The collision terms and the self-energies
of the constraint~(\ref{CE:gR}--\ref{CE:gN+})
and kinetic equations~(\ref{KE:gR}--\ref{KE:gN+}) can be easily reconstructed
from Eqs.~(\ref{Wigner-space:fermionic_eom}--\ref{Cpsi}) 
by using the methods developed in 
Refs.~\cite{ProkopecSchmidtWeinstock:2003,ProkopecSchmidtWeinstock:2004},
 and shall be addressed elsewhere.
The kinetic and constraint equations represent an 
exact tree-level description of fermionic dynamics in the presence of 
bubble walls with planar symmetry. 

We will now show how to solve equations~(\ref{CE:gR}--\ref{CE:gN+})
and~(\ref{KE:gR}--\ref{KE:gN+}) in a gradient expansion, 
but without performing flavor rotations before we decouple 
partially the equations,
in contrast to what was done in
 Refs.~\cite{ProkopecSchmidtWeinstock:2003,ProkopecSchmidtWeinstock:2004}.

\section{Constraint equations\label{sec:constraint}}

To get an idea about the classical quasiparticle limit of our solutions,
we first consider the constraint equations to lowest (classical) order, 
which are obtained from~(\ref{CE:gR}--\ref{CE:gN+}) by taking the limit
$\hat E\rightarrow 1$ and $\hat E^\dagger\rightarrow 1$, 
\bea
2(k_0 - sk_z) g_R - m\dd g_N - g_N\dd m &=& 0
\label{CR}
 \\
2(k_0 + sk_z) g_L - m g_N\dd - g_N m\dd &=& 0
\label{CL} 
\\
2k_0 g_N - is \partial_z g_N - m g_R - g_L m &=& 0
\,,
\label{CN} 
\eea
where for notational simplicity here and in the subsequent text we drop
the superscript spin index $s$.
In order to solve these equations to lowest order, 
it is convenient to make use of the self-consistency of the system of 
equations~(\ref{CE:gR}--\ref{CE:gN+}) 
and~(\ref{KE:gR}--\ref{KE:gN+}) 
({\it cf.} Ref.~\cite{ProkopecSchmidtWeinstock:2003}), and use
the solution of the kinetic equations~(\ref{KE:gN}--\ref{KE:gN+})
to lowest order, and working in the stationary limit,
in which $g_R=g_R(k^\mu,z)$, $g_{L}=g_L(k^\mu,z)$ and $g_{N}=g_N(k^\mu,z)$,
and where $\partial_t$ and $\nabla_\|$ derivatives vanish, 
\begin{eqnarray}
  g_N &=& \frac{1}{2sk_z} \Big(
                                  mg_R - g_L m 
                            \Big)
\label{KE:gN:low}
\\
  {g_N}^\dagger &=& \frac{1}{2sk_z} \Big(
                                          g_R m^\dagger - m^\dagger g_L
                                      \Big)
\,.
\label{KE:gN+:low}
\end{eqnarray}
When these equations are inserted in~(\ref{CR}--\ref{CL}), one gets 
\bea
\Big(\tilde k_0 - sk_z - \frac{1}{4sk_z}\{m^\dagger m,\cdot\}\Big) g_R
            + \frac{1}{2sk_z} m^\dagger g_L m &=& 0 
\label{CE:gR:1}
\\
\Big(\tilde k_0 + sk_z + \frac{1}{4sk_z}\{m m^\dagger,\cdot\}\Big) g_L
            - \frac{1}{2sk_z} m g_R m^\dagger &=& 0 
\,,
\label{CE:gL:1}
\eea
where $\{a,b\} \equiv ab+ba$ denotes the anticommutator.
These equations can be decoupled by multiplying~(\ref{CE:gL:1}) 
by $m$ from the right and by $m^\dagger$ from the left, and 
insering the solution into~(\ref{CE:gR:1}) (and performing an analogous 
procedure for the other equation). The result is 
\begin{eqnarray}  
\Big(k^2 - \frac{1}{2}\{m^\dagger m,\cdot\} 
        - \frac{1}{16k_z^2}\Big[m^\dagger m,[m^\dagger m,\cdot]\Big]\Big) g_R
      &=& 0 
\label{CE:gR:2}
\\
\Big(k^2 - \frac{1}{2}\{m m^\dagger,\cdot\} 
        - \frac{1}{16k_z^2}\Big[m m^\dagger,[m m^\dagger,\cdot]\Big]\Big) g_L
      &=& 0 
\,,
\label{CE:gL:2}
\end{eqnarray}
where $k^2 = \tilde k_0^2 - k_z^2 = k_0^2 - \vec k^2$, and we made use of 
%
 $
 \big\{a,\{a,f\}\big\} - 4 a f a = \big[a,[a,f]\big] 
 $.
%
These constraint equations are easily solved by transforming to 
the diagonal basis, that is 
by applying on the first (second) equation $V$ (U) from the left and 
$V^\dagger$ ($U^\dagger$) from the right, we find that 
the mass shells of $g_R$ and $g_L$ are identical,
\begin{eqnarray}
 \Big(k^2 - \frac 12(m_i^2 + m_j^2) - \frac{1}{16 k_z^2}(m_i^2-m_j^2)^2\Big)
         ({g_{R/L}}_d)_{ij} = 0
\end{eqnarray}
where $m_i^2$ ($i=1,..,N$) are the (diagonal) entries of the matrix 
$m_d^2 \equiv {\rm diag}(m_1^2,m_2^2,..,m_N^2) =
   Vm^\dagger m V^\dagger =  Um m^\dagger U^\dagger$.
The solution is given by the spectral form, 
\begin{eqnarray}
 (Vg_R^{s<}V^\dagger)_{ij} &=&  2\pi (\tilde k_0 + sk_z){\rm sign}(k_0)
 \delta \Big(
             k^2 - \frac 12(m_i^2 + m_j^2) - \frac{1}{16 k_z^2}(m_i^2-m_j^2)^2
        \Big) n_{ij}^s(k,x) 
\label{spectral-sol:gR}
\\
 (Ug_L^{s<}U^\dagger)_{ij} &=&  2\pi (\tilde k_0 - sk_z){\rm sign}(k_0)
 \delta \Big(
             k^2 - \frac 12(m_i^2 + m_j^2) - \frac{1}{16 k_z^2}(m_i^2-m_j^2)^2
        \Big) n_{ij}^s(k,x) 
\,,\quad
\label{spectral-sol:gL}
\end{eqnarray}
where $n^s_{ij}=n^s_{ij}(k,x)$ denote distribution functions.
(The indices $s$ and $<$ are here restored and 
$\tilde k_0 = {\rm sign}(k_0)(k_0^2-k_x^2-k_y^2)^{1/2}$.)
It now immediately follows that the dispersion relations
for the densities $n_{ij}^s$  are given by 
\begin{equation}
 \pm k_0 \equiv \omega_{ij} = \Big(\vec k^2 
                                  + \frac 12(m_i^2 + m_j^2) 
                                  + \frac{1}{16 k_z^2}(m_i^2-m_j^2)^2
                              \Big)^{1/2}
\label{dispersion-relation:ij}
\end{equation}
 From~(\ref{spectral-sol:gR}--\ref{spectral-sol:gL})
one infers that, while the diagonal densities $n_{ii}^s$ are projected
on the standard classical shells, $\omega_{ii} = (\vec k^2 + m_i^2)^{1/2}$,
the shells of the off-diagonal densities $n_{ij}^s$ ($i\neq j$) are
given by $\omega_{ij}$ in~(\ref{dispersion-relation:ij}),
and are in principle different for each choice of $i,j$. 
 
 A particularly simple case is when there are only two fermionic flavors
and only one off-diagonal shell. In this case $g_L$ and $g_R$ 
can be decomposed into diagonal and transverse densities as follows, 
\begin{eqnarray}
  g_R = g_R^{D} + g_R^{T}  
\,,\qquad 
  g_L = g_L^{D} + g_L^{T}  
\,,
\label{gRL:D+T}
\end{eqnarray}
where 
\begin{eqnarray}
   g_{R}^{D} = P^D_{R} g_{R} 
\,,\qquad
   g_{L}^{D} = P^D_{L} g_{L} 
\,,\qquad
   g_{R}^{T} = P^T_{R} g_{R} 
\,,\qquad
   g_{L}^{T} = P^T_{L} g_{L} 
\,.
\label{gRL:DT}
\end{eqnarray}
The projection operators are defined as 
\bea
P_R^T &=& \frac{1}{4(\mathbf{m^2})^2} \big[m^\dagger m,[m^\dagger m,\cdot]\big]
\,,\qquad
  P_R^D = 1 - P_R^T 
\label{P_R}
\\
P_L^T &=& \frac{1}{4(\mathbf{m^2})^2} \big[m m^\dagger,[m m^\dagger,\cdot]\big]
\,,\qquad
  P_L^D = 1 - P_L^T 
\label{P_L}
\eea
where
$(\mathbf{m^2})^2  = [{\rm tr}(m^\dagger m)/2]^2 - {\rm det}(m^\dagger m)
= [{\rm tr}(m m^\dagger)/2]^2 - {\rm det}(m m^\dagger)$. 
This notation has its origin in rewriting
 $mm^\dagger = \sum_{a=0}^3 (mm^\dagger)^a \sigma^a$
in terms of the Pauli algebra, 
$\sigma^a = (\mathbbm{1},\sigma^i)$, 
$[\sigma^i,\sigma^j] = 2i\epsilon^{ijl}\sigma^l$, such that 
$\mathbf{m^2} \equiv \sum_{i=1}^3 (mm^\dagger)^i 
                   = \sum_{i=1}^3 (m^\dagger m)^i$,
where we have used the fact that $\mathbf{m^2}$ is an invariant. 
In the frame in which $m^\dagger m$  ($m m^\dagger$ ) is purely diagonal, 
$g^{D}_{R}$ ($g^{D}_L$) are diagonal, and $g^{T}_{R}$ ($g^{T}_L$)
are off-diagonal and thus transverse,
which explains the notation in~(\ref{gRL:D+T}--\ref{gRL:DT}). 

Using the projectors $P^{D}_{R/L}$ and $P^{T}_{R/L}$, the constraints for
the diagonal and transversal parts decouple and read
\bea
  (k^2 - mm^\dagger ) g^{D}_L &=& 0 \\
\Big(k^2 - (mm^\dagger)^0 
   - \frac {(\mathbf m^2)^2}{4k_z^2}  \Big) g^{T}_L &=& 0 
\eea
In deriving these equations we used, $[mm^\dagger,g^{sD}_L]=0$,
$\{mm^\dagger,g^{T}_L\} = {\rm tr}(mm^\dagger)g^{T}_L
 = 2(mm^\dagger)^0g^{T}_L$ and 
$\big[mm^\dagger,[mm^\dagger,g^{T}_L]\big] = 4\mathbf{m^2}g^{T}_L$.
Similarly, we have 
\bea
  (k^2 - m^\dagger m ) g^{D}_{R} &=& 0 \\
\Big(k^2 - (m^\dagger m)^0 
   - \frac {(\mathbf m^2)^2}{4k_z^2}  \Big) g^{T}_R &=& 0 
\eea
Since both $(m^\dagger m)$ and $mm^\dagger$ are hermitean, 
$(m^\dagger m)^0=(mm^\dagger)^0$, and the dispersion relations
for the $L$- and $R$- chiralities are identical at the leading order
in gradients,
\begin{eqnarray}
  \omega_{i}^D \equiv \omega_{iL}^D = \omega_{iR}^D
                = \big(\vec k^2 + (Umm^\dagger U^\dagger)_{ii}\big)^{1/2}    
\label{omega:0DLR}
\\
  \omega^T \equiv \omega_{L}^T = \omega_{R}^T
                = \Big(\vec k^2 + (mm^\dagger)^0
                + \frac{(\mathbf{m}^2)^2}{4k_z^2}\Big)^{1/2}    
\label{omega:0TLR}
\end{eqnarray}
 An analysis of the constraint equations~(\ref{CE:gR}--\ref{CE:gN+})
shows that at higher order in gradients the diagonal and transverse shells 
mix in a manner which includes the derivative $\partial_{k_z}$, leading
to nonalgebraic constraints for the Wightman functions, and thus 
seemingly breaking the quasiparticle picture of the plasma, which questions
the validity of any on-shell description of the dynamics 
of CP-violating densities, which necessarily involve higher order gradients.
The situation is more complex however, than this simple argument seems 
to indicate. As we show in the next section, in spite of this problem
with the constraint equations, one can solve the tree-level kinetic equations
to an arbitrary high order in gradients, thanks to the fact that,
in stationary situations, the 
kinetic equations~(\ref{KE:gR}--\ref{KE:gN+}) do not
involve $k_0$, and thus the tree-level dynamics of the Wightman functions
$g_{R}$, $g_{L}$ and $g_{N}$ and the corresponding on-shell densities
(obtained by $k_0$-integration) are identical.
(In nonstationary situations off-shell effects may be important however,
which is indicated by the $k_0$ dependences appearing in the $\partial_t$ 
and $\nabla_\|$ derivatives in Eqs.~(\ref{KE:gR}--\ref{KE:gN+}).)
This also means that stationary tree-level dynamics is completely
specified by the on-shell solution of the corresponding Dirac equation,
which is by no means true in general situations.

The importance of the leading order analysis of the constraint equations
presented here stems from the fact that it allows for the on-shell projection
of the collision term and self-energies at leading order in gradients, such
that it is essential for a self-consistent derivation of the kinetic
equations for mixing fermions, provided one approximates the collision term
at leading order in gradients.

\section{Kinetic Equations to Lowest Order}

Using (\ref{KE:gN:low}--\ref{KE:gN+:low}) 
in (\ref{KE:gR}) and (\ref{KE:gL}) to lowest order, 
and working in the stationary limit, we get
\bea
     \partial_z g_R + \frac{i}{2 k_z}\com{m^\dagger m}{g_R}  &=& 0 
\\
     \partial_z g_L + \frac{i}{2 k_z}\com{m m^\dagger}{g_L} &=& 0
\,.  
\label{KE:RL:leading}
\eea
From the solutions of these equations
\begin{eqnarray}
 g_{R}(k_z,z)
   &\simeq&  {\rm exp}\Big(
                 -\frac{i}{2k_z}\int_0^z dz^\prime[m^\dagger m(z^\prime),\cdot]
                      \Big) g_{R}(k_z,0)
\,,\qquad 
\label{gR:leading}
\\
 g_{L}(k_z,z)
   &\simeq&  {\rm exp}\Big(
                 -\frac{i}{2k_z}\int_0^z dz^\prime[m m^\dagger(z^\prime),\cdot]
                      \Big) g_{L}(k_z,0)
\label{gL:leading}
\end{eqnarray}
we see that the diagonal and off-diagonal densities 
(when viewed in the diagonal basis, 
$g_{Rd} = Vg_R V^\dagger$, $g_{Ld} = Ug_L U^\dagger$)
exhibit a qualitatively different behavior.
The diagonal densities do not evolve, while 
the off-diagonals exhibit the vacuum oscillations,
well known from the neutrino studies. 
Note the identical evolution of the $L$ and $R$ chiralities, 
when viewed in the diagonal basis. 
Note further that in the case of two mixing fermions, we have 
$[m^\dagger m,g_R^D] = 0$, $[m m^\dagger,g_L^D] = 0$,
$[m^\dagger m,g_R^T] = 2i (\mathbf{m}^2\times\mathbf{g}_R^T)\cdot\vec\sigma$
and $[mm^\dagger,g_L^T]= 2i(\mathbf{m}^2\times\mathbf{g}_L^T)\cdot\vec\sigma$,
such that the transverse densities rotate with the frequency 
specified by $\vec{\mathbf{\omega}}=\mathbf{m^2}/k_z$.

\section{Kinetic Equations to Second Order \label{sec:k2o}}

 In stationary situations, one can rewrite the system of kinetic 
equations~(\ref{KE:gR}--\ref{KE:gN+})
in terms of the chiral densities $g_R$ and and $g_L$ only, valid 
for all orders in gradient expansion,
\begin{eqnarray}
 && \partial_z g_R 
      + \frac{i}{2}\Big(
                        m^\dagger\hat E \frac{1}{k_z}\big(m \hat Eg_R\big)
                   \Big)
      - \frac{i}{2}\Big( \frac {1}{k_z}\big(g_R\hat E^\dagger m^\dagger\big)
                          \hat E^\dagger m
                   \Big)
\nonumber\\
   &&\hskip 5.1cm
  - \frac{i}{2}\Big(
                    m^\dagger\hat E\frac{1}{k_z}\big(g_L \hat E^\dagger m\big)
               \Big)
+ \frac{i}{2} \Big(
                  \frac {1}{k_z}\big(m^\dagger\hat E g_L\big)\hat E^\dagger m
              \Big)
     = 0
\label{KE:gR:exact}
\\
 && \partial_z g_L
      + \frac{i}{2}\Big(
                        m\hat E\frac{1}{k_z}\big(m^\dagger\hat Eg_L\big)
                   \Big)
      - \frac{i}{2}\Big(\frac {1}{k_z} (g_L\hat E^\dagger m)
                          \hat E^\dagger m^\dagger
                   \Big)
\nonumber\\
   &&\hskip 5.1cm 
    - \frac{i}{2}\Big(
                   m\hat E\frac{1}{k_z}\big(g_R \hat E^\dagger m^\dagger\big)
                 \Big)            
 +\frac{i}{2} \Big( 
                   \frac {1}{k_z}\big(m\hat E g_R\big)
                      \hat E^\dagger m^\dagger
              \Big)
     = 0
\,.
\label{KE:gL:exact}
\end{eqnarray}
Note first that the chiral densities $g_R$ and $g_L$ couple through
derivative terms only, which justifies the use of the chiral densities
in writing the kinetic equations for mixing fermions.
Next, equations~(\ref{KE:gR:exact}) and~(\ref{KE:gL:exact}) are transformed 
into each other by the following replacements, $R\leftrightarrow L$,
$m\leftrightarrow m^\dagger$ and $s\leftrightarrow -s$ (see {\it e.g.} 
Eqs.~(\ref{CE:gR:1}--\ref{CE:gL:1})),
defining thus the symmetry, which relates the dynamics of the chiral densities
$g_R^s$ to $g_L^s$.
Furthermore, we have arrived at Eqs.~(\ref{KE:gR:exact}--\ref{KE:gL:exact})
without using the constraint equations~(\ref{CE:gR}--\ref{CE:gN+}).
This procedure has the advantage that $k_0$ appears
nowhere in Eqs.~(\ref{KE:gR:exact}--\ref{KE:gL:exact}),
implying that the kinetic equations for 
the distribution functions $f_{Rs\pm}$ and $f_{Ls\pm}$, 
defined as the (positive and negative) frequency
integrals of $g_{R}^s$ and  $g_{L}^s$,
have exactly the same form as~(\ref{KE:gR:exact}--\ref{KE:gL:exact}), 
resolving thus the problem of closure of the on-shell kinetic equations. 
We emphasize that the (tree-level) closure is thus achieved, even though the 
constraint equations are nonalgebraic. 
One consequence of the nonalgebraic nature of the constraint equations
is a coupling between the off-diagonal and diagonal densities, which 
is nevertheless implemented in a self-consistent manner
into the kinetic equations~(\ref{KE:gR:exact}--\ref{KE:gL:exact})
(through the higher derivative terms), without ever referring to 
the on-shell structure of the system. 
If one had attempted to further decouple the equations for $g_R$ and $g_L$, 
one would have found out that this could be achieved 
by making use of the constraint equations~(\ref{CE:gR}--\ref{CE:gN+}),
which would reintroduce the dependences on $k_0$ and $s$, which 
is not explicit in equations~(\ref{KE:gR:exact}--\ref{KE:gL:exact}).

Upon expanding $\hat E$ and $\hat E^\dagger$ in~(\ref{EandEdagger})
to second order in gradients, 
\begin{eqnarray}
 \hat E &=& 1 
         + \frac i2 \overset{\leftarrow}{\partial}_z
            \overset{\rightarrow}{\partial}_{k_z}
         - \frac 18 (\,\overset{\leftarrow}{\partial}_z
            \overset{\rightarrow}{\partial}_{k_z})^2
         + ..
\nonumber\\
\hat E^\dagger &=& 1
         - \frac i2 \overset{\leftarrow}{\partial}_{k_z}
            \overset{\rightarrow}{\partial}_z
         - \frac 18 (\,\overset{\leftarrow}{\partial}_{k_z}
            \overset{\rightarrow}{\partial}_z)^2
         + ..
\label{hatE:second-order}
\end{eqnarray}
we can write the chiral kinetic 
equations~(\ref{KE:gR:exact}--\ref{KE:gL:exact}),
truncated at second order in gradients as follows, 
\begin{eqnarray}
 &&\!\!\!\!\!\!
   k_z\partial_z g_R 
 + \frac{i}{2}\big[m^\dagger m,g_R\big]
\nn \\
 &&\hskip 0.9cm 
 - \frac{1}{4}\big\{(m^\dagger m)^\prime,\partial_{k_z}g_R\big\}
 + \frac{1}{4k_z}\big(
                     {m^\dagger}^\prime {m} g_R + g_R m^\dagger {m}^\prime
                   \big)
 - \frac{1}{4k_z}\big(
                     {m^\dagger}^\prime g_L {m}  +  m^\dagger g_L {m}^\prime
                   \big)
\nn\\
 &&\hskip 0.9cm 
 - \frac{i}{16}\big[(m^\dagger m)^{\prime\prime},\partial_{k_z}^2g_R\big]
 + \frac{i}{8k_z}
      \big[
          {m^\dagger}^{\prime}{m}^\prime,\partial_{k_z}g_R
      \big]
\nn \\
 &&\hskip 0.9cm 
 + \frac{i}{8}
     \Big(
          {m^\dagger}^{\prime\prime}{m}\partial_{k_z}\big(\frac{g_R}{k_z}\big)
    \!-\! \partial_{k_z}\big(\frac{g_R}{k_z}\big){m^\dagger}{m}^{\prime\prime}
     \Big)
 - \frac{i}{8}
     \Big(
          {m^\dagger}^{\prime\prime}\partial_{k_z}\big(\frac{g_L}{k_z}\big){m}
     \!-\! {m^\dagger}\partial_{k_z}\big(\frac{g_L}{k_z}\big){m}^{\prime\prime}
     \Big)
\label{gR:2nd-order:2o}
\simeq 0
\,.
\qquad
\label{gR:2nd-order}
\end{eqnarray}
The kinetic equation for $g_L$ is obtained from~(\ref{gR:2nd-order})
simply by exacting the replacements, 
$g_R\leftrightarrow g_L$ and $m\leftrightarrow m^\dagger$. 

 To get a rough idea on what are the criteria for the applicability
of the gradient expansion, let us first 
recall the relevant criterion for the one fermion case, 
\begin{equation}
  \hbar\partial^\mu \ll \hbar k^\mu
\,.
\label{grad-expansion:validity:1}
\end{equation}
Since $\partial_z \sim 1/L_w$, this criterion was used to coin  
the term ``thick wall regime'' in baryogenesis studies. The proper
interpretation of~(\ref{grad-expansion:validity:1}) is closely related
to the validity of the WKB approximation in quantum mechanics, 
which is valid when the de Broglie wavelength of the excitations is small
in comparison to the region over which the background (mass) varies.
In the case of mixing fermions however,
an additional complication arises from the evolution 
of the off-diagonal densities. Indeed, from the leading 
order solution~(\ref{gL:leading}) we find, 
 $\partial_{k_z} (g^T_{R,L})_{ij}
   \sim  - i\big[\int_0^z dz'(m_i^2-m_j^2)/(2k_z^2)\big] (g^T_{R,L})_{ij}$, 
such that the dynamics of the transverse densities can jeopardize 
the validity of the gradient expansion. Indeed, 
the gradient expansion applies provided formally,
$\big[\int_0^z dz'(m_i^2-m_j^2)/(2k_z^2)\big] \partial_z \ll 1$ 
($\forall i,j$) is satisfied,
such that the dependence on $L_w \sim 1/\partial_z \sim \int_0^z dz'$ 
roughly cancels out. This implies an additional
and qualitatively new criterion for the validity of the gradient expansion, 
$m_i^2-m_j^2 \ll k_z^2 \, (\forall i,j)$. 

This criterion should be taken with great caution, however. 
Being derived without any reference to decoherence of
off-diagonal densities,
this criterion may be in many situations too stringent.
In particular, large differences in mass eigenvalues and low 
momenta $k_z$ imply fast oscillations of transverse densities,
which are more prone to decoherence by rescatterings, and thus destruction,
than slowly oscillating densities.
We thus conclude that
a complete analysis of applicability of the gradient expansion
for mixing fermions is at the moment not available.

\subsection{Reduction to the diagonal limit}
\label{Reduction to the diagonal limit}

 Let us now consider the diagonal part of the 
kinetic equation~(\ref{gR:2nd-order}) and its left-handed counterpart, 
\begin{eqnarray}
 &&k_z \partial_z g_R
   -\frac14\big\{{(m^\dagger m)^\prime}^D\!\!,\partial_{k_z}g_R\big\}
 +\frac{1}{8k_z}\big\{{(m^\dagger m)^\prime}^D\!\!,g_R-\hat g_L\big\}
\nonumber\\
&& \hskip 6cm
  - \frac{i}{16}\Big\{
                   {({m^\dagger}^\prime m-{m^\dagger} m^\prime)^\prime}^D\!\!,
                       \partial_{k_z}\frac{g_R-\hat g_L}{k_z}
                          \Big\}
 =0 
\qquad
\label{gR2}
\\
 &&k_z \partial_z g_L
   -\frac14\big\{{(m m^\dagger)^\prime}^D\!\!,\partial_{k_z}g_L\big\}
 +\frac{1}{8k_z}\big\{{(m m^\dagger)^\prime}^D\!\!,g_L-\hat g_R\big\}
\nonumber\\
&& \hskip 6cm
 - \frac{i}{16}\Big\{
                 {({m}^\prime m^\dagger-{m} {m^\dagger}^\prime)^\prime}^D\!\!,
                       \partial_{k_z}\frac{g_L-\hat g_R}{k_z}
                          \Big\}
 = 0
\,,
\qquad
\label{gL2}
\end{eqnarray}
where $g_{Rd} =Vg_RV^\dagger\equiv U\hat g_RU^\dagger$
 and $g_{Ld} = Ug_LU^\dagger \equiv V\hat g_LV^\dagger$
 are assumed to be diagonal. Since in the diagonal basis
 $\big(V(m^\dagger m)^\prime V^\dagger\big)^D
   = ({m_d^2})^\prime = \big(U(mm^\dagger)^\prime U^\dagger\big)^D$,
and 
\begin{equation}
  \big(
    V({m^\dagger}^{\prime\prime}m-{m^\dagger}m^{\prime\prime})V^\dagger
  \big)^D
= \big(
    U(m{m^\dagger}^{\prime\prime}-m^{\prime\prime}{m^\dagger})U^\dagger
  \big)^D
\,,
\label{SR-SL:identity}
\end{equation}
we see that the equations for 
$g_{0d} \equiv (g_{Rd}-g_{Ld})/2$ and 
$g_{3d} \equiv (g_{Rd}+g_{Ld})/2$ decouple,  
\begin{eqnarray}
k_z \partial_z g_{0d}
   -\frac12(m_d^2)^\prime\partial_{k_z}g_{0d}
  -  \frac{is}{4\tilde k_0}
    \big(V({m^\dagger}^\prime m-{m^\dagger} m^\prime)^\prime V^\dagger\big)^D
                       \partial_{k_z}g_{0d}
 &=& 0
\label{g0(2)}
\\
\partial_z g_{3d}
   -\frac{1}{2}(m_d^2)^\prime\partial_{k_z}\frac{g_{3d}}{k_z}
   &=&0
\,.
\label{g3(2)}
\end{eqnarray}
Note that the leading order solution, 
$g_{3d}^{(0)} = (sk_z/\tilde k_0)g_{0d}^{(0)}$ 
solves~(\ref{g3(2)}), where $g_{0d}^{(0)}$ is given
 in~(\ref{g0d:leading order}), such that up to second order in 
gradients there is no source for the axial density in the 
diagonal approximation.

 On the other hand, the form of the vector equation~(\ref{g0(2)})
suggests that in the static case (\ref{g0(2)}) can be solved 
exactly. This is in fact not quite so, since the quantities 
in~(\ref{SR-SL:identity}) are not a total derivative. Nevertheless, 
an approximate solution can be found for a static 
wall~\cite{ProkopecSchmidtWeinstock:2003, ProkopecSchmidtWeinstock:2004}:
\bea
g_{0d} \simeq 2\pi|\tilde k_0|\, 
  \delta\Big(k^2 - |m_d|^2 
 + \frac{is}{2\tilde k_0}
         \big(
              V({m^\dagger}^\prime m - m^\dagger m^\prime)V^\dagger
         \big)^D\Big) n_0(k_0)
\,,\quad n_0 = \frac{1}{{\rm e}^{k_0/T_c}+1}
\,.\qquad 
\label{exact solution:diagonal}
\eea
This is a good approximation when the mass matrix is approximately diagonal, 
or when the transverse elements of $V{V^\dagger}^\prime$ and 
$U{U^\dagger}^\prime$ are small. Eq.~(\ref{g0(2)})
reproduces the result first derived in 
Refs.~\cite{KainulainenProkopecSchmidtWeinstock:2001,
KainulainenProkopecSchmidtWeinstock:2002}, where the last term
sources a CP-violating current, of crucial importance for 
electroweak scale baryogenesis studies. 
  In sections~\ref{Domination by diagonal parts},
\ref{Local Contributions to the Currents in the MSSM} 
and~\ref{Numerical Results of Transport in the MSSM}
 we make a quantitative comparison of the second order 
diagonal source in~(\ref{g0(2)})
 with the first order (transverse) sources, which we discuss
in detail in the next section.

\section{CP-violating sources}
\label{CP-violating sources}

Let us define CP symmetry as the following transformations
of the Dirac spinors (up to an irrelevant phase),
\begin{equation}
 \psi^{cp}(u) \equiv {\cal CP}{\psi(u)}({\cal CP})^\dagger 
        = i\gamma^2 \bar\psi^T(\bar u)
\,,\qquad
 \bar \psi^{cp}(u) \equiv {\cal CP}{\bar\psi(u)}({\cal CP})^\dagger 
        =  \psi^T(\bar u)i\gamma^2
\label{CP:psi:def}
\end{equation}
%
%
%
One finds that the kinetic
equation~(\ref{Wigner-space:fermionic_eom})
transforms as~\cite{ProkopecSchmidtWeinstock:2003},
\begin{eqnarray}
  \Big( \slash{\bar k}
          + \frac i2 \slash{\partial}_{\bar x}
          - m_h^*(x)
            {\rm e}^{-\frac{i}{2}\stackrel{\leftarrow}{\partial}_x
                      \cdot\,\partial_k}
          + i\gamma^5m_a^*(x)
            {\rm e}^{-\frac{i}{2}\stackrel{\leftarrow}{\partial}_{\!x}
                   \cdot\,\stackrel{\rightarrow}{\partial_k}}
  \Big)S^{cp<,>}(k,x)
   =     -\gamma^0\gamma^2  {\cal C}_\psi^*(-k,x)\gamma^0\gamma^2
\,,
\label{Wigner-space:fermionic_eom:cp}
\end{eqnarray}
where we neglected the self-energies. 
The Wightman functions transform as, 
\begin{equation}
  S^{<,>}(k,x) \stackrel{{\cal CP}}{\longrightarrow} 
     -\gamma^2 {S^{>,<T}(-\bar k,\bar x)} \gamma^2
        \equiv {S^{cp}}^{<,>}(k, x)
\label{S<>:cp:definition}
\end{equation}
A comparison of Eqs.~(\ref{Wigner-space:fermionic_eom}) 
and~(\ref{Wigner-space:fermionic_eom:cp}) reveals that
in the wall frame, in which $m_{h,a} = m_{h,a}(z)$, 
the CP transformation of the flow term 
is in our context equivalent to the transformations
%
\begin{eqnarray}
  m_h\rightarrow m_h^*
\,,\qquad
  m_a\rightarrow -m_a^*
\,,\qquad
 (m\rightarrow m^*,\;\; m^\dagger \rightarrow m^T)
\,,
\label{CP:effective}
\end{eqnarray}
leaving {\it e.g.} $k_z^{\pm 1}\partial_z$ 
and $\partial_z\partial_{k_z}$ invariant.
From these rules the CP transformed equation~(\ref{gR:2nd-order}) 
can be written as follows,
\begin{eqnarray}
 &&\!\!\!\!\!
   k_z\partial_z g_R^{cp} 
 + \frac{i}{2}\big[m^T m^*,g^{cp}_R\big] 
\label{gRcp:2nd-order}
 \\
 &&\hskip 0.2cm 
 - \frac{1}{4}\big\{(m^T m^*)^\prime,\partial_{k_z}g_R^{cp}\big\}
 + \frac{1}{4k_z}\big(
                      {m^T}^\prime {m^*} g_R^{cp} 
                    + g_R^{cp} m^T {m^*}^\prime
                   \big)
 - \frac{1}{4k_z}\big(
                      {m^T}^\prime g_L^{cp} {m^*}
                    +  m^T g_L^{cp} {m^*}^\prime
                   \big)
\nn \\
 &&\hskip 0.2cm 
 - \frac{i}{16}\big[(m^T m^*)^{\prime\prime},\partial_{k_z}^2g_R\big]
 + \frac{i}{8k_z}
      \big[
          {m^T}^{\prime}{m^*}^\prime,\partial_{k_z}g_R
      \big]
\nn \\
 &&\hskip 0.2cm 
 + \frac{i}{8}
      \Big(
          {m^T}^{\prime\prime}{m^*}\partial_{k_z}\big(\frac{g_R}{k_z}\big)
        - \partial_{k_z}\big(\frac{g_R}{k_z}\big){m^T}{m^*}^{\prime\prime}
      \Big)
 - \frac{i}{8}
      \Big(
          {m^T}^{\prime\prime}\partial_{k_z}\big(\frac{g_L}{k_z}\big){m^*}
        - {m^T}\partial_{k_z}\big(\frac{g_L}{k_z}\big){m^*}^{\prime\prime}
      \Big)
\label{gRcp:2nd-order:2o}
 \simeq 0
\nn
\,.
\end{eqnarray}
Our primary goal is to identify the CP-violating sources in the system.
A na\"\i ve way of doing that would be to 
subtract Eq.~(\ref{gRcp:2nd-order}) from Eq.~(\ref{gR:2nd-order}),
and identify the terms in the equation for
$\delta g^{cp}_R = g_R-g_R^{cp}$ which involve a CP-violating operator 
acting on $\bar g_R = (g_R+g_R^{cp})/2$, thus representing 
mixing of CP-odd and CP-even densities. This procedure leads to 
equations with indefinite transformation properties under flavor rotations,
which is a consequence of the indefinite transformation properties
of the newly defined densities $\delta g_R^{cp}$ and $\bar g_R$, making it
difficult to disentangle the genuine CP-violating densities
from the apparent, but possibly spurious, CP-violating densities. 

%
%

In the following, we propose a method, which allows to extract the CP 
violation from the kinetic equations~(\ref{gR:2nd-order})
and~(\ref{gRcp:2nd-order}). 
We see from Eq.~(\ref{S<>:cp:definition}) not only
that CP symmetry is broken by the theory, but also that it depends on the 
basis in which the usual definition of CP conjugation is used. Therefore 
it is helpful, instead of the usual CP conjugation,
to introduce another operation
\bea 
\Q \, g(k,x) \, \Qd =  \CP \, g^T(k,x) \, \CPd = g(-\bar k, \bar x ) 
.
\eea   
For the second equality, the relation (\ref{S<>:cp:definition}) has been used.

Note that we do not want to undo a part of the usual CP conjugation, but just 
want to include an additional sign change of the imaginary elements
in flavor space (the coefficients of $\sigma_2$ in the case of two
mixing fermions). This operator has some 
nice properties, that are absent in the case of the standard CP conjugation.
For example, Q commutes with basis transformations, implying that 
if a quantity transform in a definite manner under Q conjugation 
in the mass eigenbasis, it will transform in the same way in the flavor basis.
(This can be easily checked by noting that
 $\CP \, V \, \CPd = V^*, \CP \, U \, \CPd = U^*$.)
In addition, Q conjugation agrees with the usual CP conjugation for 
the remaining coefficients in flavor space
 (of ${\mathbf 1},\sigma_1$ and $\sigma_3$ in the two flavor case);
 in particular it agrees for the diagonal elements. 
Hence, in order to generate a CP-violating effect, 
one has to create Q-odd terms at least in the off-diagonal terms
 in the mass eigenbasis, such that CP violation becomes manifest in the 
diagonal terms of the flavor basis. 

Knowing this, we look for the transformation properties of the 
kinetic equation under the Q conjugation and observe that all terms 
that come from an even order of the gradient expansion acquire an additional
minus sign relative to the odd contributions from the 
gradient expansion. As a consequence, in the one flavor case one has
to take second order terms into account in order to break Q
(in this case Q is of course equivalent to CP),
since there are no zeroth order terms. This leads to semiclassical force 
induced baryogenesis.

In the multi flavor case, an expansion up to first order is 
sufficient as long as the zeroth order contributes 
(otherwise the Green function is 
everywhere diagonal in the mass eigenbasis and the problem 
reduces to the one flavor case as discussed in
 \cite{ProkopecSchmidtWeinstock:2003,ProkopecSchmidtWeinstock:2004}).

Let us recall the kinetic equation of the right handed density to 
first order
\bea
 &&\!\!\!\!\!\!
   k_z\partial_z g_R 
 + \frac{i}{2}\big[m^\dagger m,g_R\big] 
 - \frac{1}{4}\big\{(m^\dagger m)^\prime,\partial_{k_z}g_R\big\} \nn \\
 &&\hskip 0.9cm 
 + \frac{1}{4k_z}\big(
                     {m^\dagger}^\prime {m} g_R + g_R m^\dagger {m}^\prime
                   \big)
 - \frac{1}{4k_z}\big(
                     {m^\dagger}^\prime g_L {m}  +  m^\dagger g_L {m}^\prime
                   \big) \simeq 0.
\label{KE:R:1sr-order}
\eea

The Q-conjugate equation is 
\bea
 &&\!\!\!\!\!\!
   k_z\partial_z g^{Q}_R 
 - \frac{i}{2}\big[m^\dagger m,g^{Q}_R\big] 
 - \frac{1}{4}\big\{(m^\dagger m)^\prime,\partial_{k_z}g^{Q}_R\big\} \nn \\
 &&\hskip 0.9cm 
 + \frac{1}{4k_z}\big(
                     {m^\dagger}^\prime {m} g^{Q}_R + g^{Q}_R m^\dagger {m}^\prime
                   \big)
 - \frac{1}{4k_z}\big(
                     {m^\dagger}^\prime g^{Q}_L {m}  +  m^\dagger g^{Q}_L {m}^\prime
                   \big) \simeq 0.
\eea
such that only the sign of the second (zeroth order commutator) 
term is affected.

To solve this equation, we will first determine the lowest 
order solution and then expand around it. The best way to determine it
is in the mass eigenbasis, since in this basis the 'direction' of the mass in
flavor is fixed. In the flavor basis we can mimic this property by adding 
a term that explicitly compensates for the $z$-dependent basis transformation.  
In the case of a static wall profile,
we can in addition include the diagonal part of the 
third term, since it belongs to the classical Boltzmann-like flow, and  
we know how to handle it from the one flavor case.
Diagonal means in this context that it commutes with $m^\dagger m$ as it 
is indicated by our notation introduced in section \ref{sec:constraint}.
Our lowest order solution fulfills 
\bea
 &&\!\!\!\!\!\!
   k_z\partial_z g^{(0)}_R 
 + \frac{i}{2}\big[m^\dagger m,g^{(0)}_R\big] 
 - \frac{1}{4}\big\{(m^\dagger m)^\prime,\partial_{k_z} g^{(0)}_R\big\}^D 
 - k_z\com{{V^\dagger}^\prime V}{g_R^{(0)}} = 0
\,,
\label{kin:left}
\eea
where the first commutator term vanishes, since there is no 
source for the transversal parts. Now since
 $V\big\{(m^\dagger m)^\prime,\partial_{k_z} g^{(0)}_R\big\}^D V^\dagger
   = 2 (m_d^2)^\prime\partial_{k_z} g^{(0)}_{Rd}$,
the solution of~(\ref{kin:left}) is simply, 
\begin{equation}
 g_R^{(0)} = \Big(1+\frac{sk_z}{\tilde k_0}\Big) 
              V^\dagger g_{0d}^{(0)} V
\,.
\end{equation}
Here $g_{0d}^{(0)}$ is the diagonal vector density in the mass-eigenbasis 
of the spectral form, 
\begin{equation}
  g_{0d}^{(0)} = 2\pi |\tilde k_0| \delta(k^2 - |m_d|^2) n
\,,
\end{equation}
and $n=n(k_0\vec k_\|)$ is a distribution function.
In thermal equilibrium, which is formally obtained in the limit of
large damping (frequent collisions), $n$ reduces to the Bose-Einstein
 distribution, $n\rightarrow n_0 = 1/[\exp(k_0/T)+1]$.
In this case the collision term vanishes
(this is also obtained by imposing the Kubo-Martin-Schwinger condition
on the Wightman functions). 
All influences of the changing background are then negligible and the
Green function depends only locally on the mass.

The transverse part of the deviation from $g_R^{(0)}$ is in 
the next order given by
\bea
 &&\!\!\!\!\!\!
   k_z\partial_z g^{T(1)}_{R} 
 + \frac{i}{2}\big[m^\dagger m, g^{T(1)}_{R}\big] + k_0 \Gamma \, g^{T(1)}_{R} 
- k_z\com{{V^\dagger}^\prime V}{g_R^{T(1)}}=  S_{R}^{(1)}
\nn \\ 
&& \;\; 
 S_{R}^{(1)} \equiv - k_z\com{{V^\dagger}^\prime V}{g_R^{(0)}}^T 
+ \frac{1}{4}\acom{(m\dd m)^\prime }{\partial_{k_z} g^{(0)}_{R} }^T
 \nn \\
 && \quad\quad \;\,
-\; \frac{1}{8k_z}
  \com{ m\ddp m - m\dd m^\prime }{ g^{(0)}_{R}- \hat g_L^{(0)}}^T
 -\frac{1}{8k_z}\acom{(m\dd m)^\prime}{g^{(0)}_{R} - \hat g_L^{(0)}}^T
,
\qquad
\label{Source:complete}
\eea
where we defined $\hat g_L^{(0)} \equiv V\dd Ug^{(0)}_{L}U\dd V$.
Here we have introduced a damping rate $\Gamma$ (which can be arbitrary small)
to fulfill boundary conditions at infinity 
($g^{(1)}_R \rightarrow 0$ for $z \rightarrow \pm \infty$). 
 At the same time this helps to cure the infrared divergencies in the sources. 
For simplicity, in this work we assume that the damping $\Gamma$  is flavor
blind, {\it i.e.} we take it to be proportional to the unity matrix
in flavor space.

The (particular) solution of equation~(\ref{Source:complete})
in the mass eigenbasis is formally given by
\bea
g_{Rd}^{T(1)} (z) = \int_{-\infty}^{+\infty} W(z,z^\prime) 
S_{Rd}^{(1)}(z^\prime) dz^\prime,
 \label{Greens:complete}
\eea
with the kernel
\bea
W(z,z^\prime) &=&  \frac{1}{k_z} 
 \big[ \theta(k_z k_0) \theta(z-z^\prime) 
 - \theta(-k_z k_0) \theta(z^\prime-z) \big] \exp \Big(
- \frac{k_0}{k_z}\Gamma (z-z^\prime) \Big)
\nn \\
&&  \times \exp \left( -\frac{i}{2k_z} \int_{z^\prime}^{z} 
   \com{m_d^2(y)}{\,\cdot\, } dy \right) 
\,.
\label{Greens:all}
\eea
The exponential function is understood as the power series
in nested commutators, and the source is rotated into the mass eigenbasis
$S_{Rd}^{(1)}=V \, S_R^{(1)} V^\dagger$.

From this we can deduce the part of $g_R$ that breaks the 
Q-symmetry ($g_R^{\slash{\!Q}} = g_R - g_R^Q$)
\bea
g_{Rd}^{\slash{Q}} (z) = \int_{-\infty}^{+\infty} W_{\slash{Q}} (z,z^\prime) 
S_{Rd}^{(1)}(z^\prime) dz^\prime
 \label{GreensQodd}
\eea
with
\bea
W_{\slash{Q}}(z,z^\prime) 
   &=&  \frac{1}{k_z}  \big[ \theta(k_z k_0) \theta(z-z^\prime) 
 - \theta(-k_z k_0) \theta(z^\prime-z) \big]
 \exp \Big( - \frac{k_0}{k_z}\Gamma (z-z^\prime) \Big)
 \nn \\
&&  \times (-i) \sin \Big( \frac{1}{2k_z} \int_{z^\prime}^{z} \com{m_d^2(y)}
{\,\cdot\, } dy \Big) 
\,.
\label{Q-breaking W}
\eea
The CP-violating diagonal part of $g_R$ in the flavor basis
is given by $ Tr \big( \sigma_3 \, V g_{Rd}^{\slash{Q}} V\dd \big)$.

\subsection{Local Sources}

Since our source is already first order in gradients, we solve the integral 
(\ref{GreensQodd}) by expanding all functions around the position $z$ and 
keep only the first Taylor coefficients. This procedure is justified provided 
 $\Gamma \, L_w \gg 1$. For the MSSM this leads to 
$\Gamma \gg L_w^{-1}\simeq {T_c}/{20}$~\cite{Moreno:1998bq}, 
with the expansion parameter
of the gradient expansion $T_c\, L_w \simeq 20$.
 Since the off diagonal entries 
are coherent superpositions of particle states, $\Gamma$ characterizes
 the inverse decoherence length. On the physical grounds we expect  $\Gamma$
 to be at least as large as the thermalization rate,
which we take to be of the order the thermalization rate for the W bosons,
 $\Gamma\sim\Gamma_W\simeq\alpha_wT_c$~\cite{MooreProkopec:1995}.
 To get its detailed form and magnitude would require a quantitative analysis
of the collision term for mixing fermions however, 
which is beyond the scope of this work.
Here we take $\Gamma$ of the order the thermalization rate
and proportional to unity in flavour space.

 Assuming that $\Gamma$ induces an efficient flavor decoupling and 
making a leading order approximation of the sine function in
Eq.~(\ref{Q-breaking W}), 
the expression contributing at leading (first) order in gradients 
acquires the following simple local form, 
\bea
g_R^{T(1)}
 &=& \Big(k_0\Gamma-\frac{i}{2}\com{m\dd m}{\,\cdot\,} \Big)^{-1} S_R^{(1)}
\,,
\eea
such that the Q-breaking part reads
\bea
g_R^{T\slash{Q}} 
= \frac{\frac{i}{2}\com{m\dd m}{\,\cdot\,}}
  {k_0^2\Gamma^2 + \frac{1}{4}\com{m\dd m}{\,\com{m\dd m}{\,\cdot\,}}} 
                 S_R^{(1)}
\,.
\label{Green:short}
\eea

Since only transverse sources lead to Q-breaking terms 
we rewrite the expression (\ref{Source:complete}) in the more 
compact form
\bea
S_R^{(1)}
 &=&  - k_z\com{{V^\dagger}^\prime V}{g_R^{(0)}}
+ \frac{1}{4}\acom{(m\dd m)^\prime }
                  {\partial_{k_z}g^{(0)}_{R}}^T
 \nn \\
 &&
- \frac{s}{4\tilde k_0}\com{ m\ddp m - m\dd m^\prime }{ \hat g^{(0)}_0}^T
 - \frac{s}{4\tilde k_0}\acom{(m\dd m)^\prime}{\hat g^{(0)}_0}^T 
\,,
\label{Source:transverse}
\eea
where $\hat g^{(0)}_3 \equiv V^\dagger g_{3d}^{(0)} V 
        = (g_R^{(0)}-V^\dagger U g^{(0)}_{L} U^\dagger V)/2$,
and we made use of the leading order constraint equation, 
$\hat g^{(0)}_3 = s (k_z/\tilde k_0) \hat g_0^{(0)}$, with
$\hat g_0^{(0)} \equiv  V^\dagger g_{0d}^{(0)} V$ and 
\begin{equation}
g_{0d}^{(0)} = 2\pi|\tilde k_0|\delta(k^2 - |m_d|^2)n
\label{g0d:leading order}
\,,
\end{equation}
where in equilibrium and for a static wall the occupation number reduces to
the Fermi-Dirac distribution, 
$n\rightarrow n_0 = ({\rm exp}(k_0/T_c)+1)^{-1}$.

Here we can establish for the first time the important fact
that in our treatment
a static wall does not induce any CP-violating charge densities.
 In the local approximation~(\ref{Green:short})
the following integrals are relevant for the calculation of the CP-violating
source current
\bea
\! j_R^{T\slash{Q}} &=&   -\frac{s}{4}\!\int\!\!\frac{d^4k}{(2\pi)^4}
 \frac{\frac{i}{2}\com{m\dd m}{\,\cdot\,}}
  {k_0^2\Gamma^2 \!+\! \frac{1}{4}\com{m\dd m}{\,\com{m\dd m}{\,\cdot\,}}} 
           \frac{k_0}{\tilde k_0^2} \nn \\
 &\times&   \bigg(\!
	4k_z^2\com{{V^\dagger}^\prime V}{\hat g^{(0)}_0}
       +  \com{ m\ddp m \!-\! m\dd m^\prime }{ \hat g^{(0)}_0}
       +  \acom{(m\dd m)^\prime}{\hat g^{(0)}_0}^{T}
    \!\bigg)
.
\label{currentsource:local}
\eea
The left-handed source current $j_L^{T\slash{Q}}$ is obtained
 simply by the substitutions,
 $s\rightarrow -s$, $m\leftrightarrow m^\dagger$ and 
$\hat g_0^{(0)} = V^\dagger g_{0d}^{(0)} V\rightarrow U^\dagger g_{0d}^{(0)}U$.
The second term in Eq.~(\ref{Source:transverse}) does not
contribute to the current~(\ref{currentsource:local}), since 
 the integral over the momenta vanishes for this term.

\subsection{Nonlocal Sources}

To evaluate (\ref{GreensQodd}) we could just solve the integral numerically.
However this would involve some technical and physical shortcomings.
First, the integrand is oscillating with a frequency
 $\omega \sim {\Lambda}/{k_z}$, which makes numerical evaluation hard.
Second, since we have parametrized the collision terms in the kinetic equation 
by just one parameter, our solution does not show the expected behavior in
certain regions of parameter space. {\it E.g.} we expect that collisions
help to isotropize the deviation from equilibrium, while the solution
to equation (\ref{GreensQodd}) has a strong $k_z$ dependence, but
almost no $k_{||}$ dependence 
($k_{||}$ denotes the momentum parallel to the wall).
Another feature which may play an important role is diffusion,
by which particles get transported typically to distances
\begin{equation}
\ell_{\rm diff} \simeq \frac{2D}{v_w+(v_w^2+4\Gamma D)^{1/2}}
\end{equation}
in front of the wall, where $D$ denotes the diffusion constant,
 $v_w$ the wall velocity, and $\Gamma$ the 
damping. In systems with small $\Gamma$ and/or
 large $D$, such that $v_w\gg 4\Gamma D$ is satisfied, 
the diffusion tail may be large, $\ell_{\rm diff}\simeq D/v_w$.
Since for charginos of the MSSM, the diffusion constant and 
the wall velocity are rather small, $D\sim 10/T$,  $v_w\leq 0.1$ and
the damping quite large, $\Gamma \sim \alpha_w T$~\cite{MooreProkopec:1995},
$v_w^2 \ll 4 \Gamma D$ is amply fulfilled,
and we can estimate the diffusion `tail' to be
$\ell_{\rm diff}\simeq (D/ \Gamma)^{1/2}\sim 15/T$. Since diffusion 
is symmetric and it extends to distances of the order the wall thickness,
we expect it to be captured reasonably well by our simple model of damping.

 To cure the shortcomings related to the local approximation,
we shall solve the kinetic equation (\ref{Source:complete}) by a fluid 
{\it Ansatz} in the mass eigenbasis
\bea
g_{Rdij}^{(1)} &=& 2\pi
  \sum_{a=0}^N T_c^{-a} \mu_{aij} (k_z - v_w k_0)^a  
\big[\partial_{k_0} n_0\big(\gamma(k_0 - v_w k_z)\big)\big]
\tilde\omega_{ij}\delta(k_0^2 - \omega_{ij}^2)
\,,
\label{fluid:ansatz}
\eea
where $\omega_{ij}$ are given by the lowest order 
on-shell conditions~(\ref{dispersion-relation:ij}),
$\tilde \omega_{ij}^2 = \omega_{ij}^2 - \vec k_\|^2$, 
and $n_0(x) = 1/[{\rm exp}(x/T_c)+1]$ 
is the Fermi-Dirac distribution function. 
Note that the fluid {\it Ansatz}~(\ref{fluid:ansatz})
captures the chiral nature of the first order 
solution, which is, for example, expressed by the 
leading order constraint relation, 
$g_{Rdij}^{(1)} = \big(1+s{k_z}/{\tilde k_0}\big)g_{0dij}^{(1)}$.

 If one now takes the first $N$ momenta of the kinetic 
equation~(\ref{Source:complete}), defined as 
$\int_{k_0>0}[d^4 k/(2\pi)^4](k_0/\tilde k_0) (k_z/T_c)^l$ $\;(l=0,..,N)$, 
one gets a matrix equation of the form (here and below 
we suppress the $i,j$ indices):
\bea
A \partial_z \mu + \frac{i}{2}B\big[m_d^2,\mu\big] 
                 + \Gamma C\mu = D
\,.
\label{momenta:matrix}
\eea
$A$, $B$, $C$ are matrices and $\mu$ and $D$ vectors in the 
$a,b$ space ($a,b\in \{0..N\}$),
with
\bea
A_{ab} &=& T_c^{-a-b} \int_{k_0>0} \frac{d^4k}{(2\pi)^3}|k_0|
   (k_z - v_w k_0)^a  k_z^{b+1}
\big[\partial_{k_0} n_0(\gamma(k_0 - v_w k_z))\big]
   \delta(k_0^2 - \omega^2)
 \\
B_{ab} &=& T_c^{-a-b} \int_{k_0>0} \frac{d^4k}{(2\pi)^3}|k_0|
 (k_z - v_w k_0)^a k_z^b
\big[\partial_{k_0} n_0(\gamma(k_0 - v_w k_z))\big]
 \delta(k_0^2 - \omega^2)
 \nn \\
C_{ab} &=& T_c^{-a-b} \int_{k_0>0} \frac{d^4k}{(2\pi)^3}|k_0|
 (k_z - v_w k_0)^a k_0 k_z^b
\big[\partial_{k_0} n_0(\gamma(k_0 - v_w k_z))\big]
 \delta(k_0^2 - \omega^2)
 \nn \\
D_{a} &=& -\frac{s}{4T_c^a} \!
 \int_{k_0>0}\! \frac{d^4k}{(2\pi)^3}\frac{k_0}{|\tilde k_0|}
                k_z^a \delta(k_0^2 - \omega^2) \nn \\
  &\times&  \bigg(\!
	 4k_z^2\com{ V V^{\dagger\prime} }{ n_0}
          + \com{ V(m\ddp m - m\dd m^\prime)V^\dagger }{ n_0}^{T}
       +  \acom{V(m\dd m)^\prime V^\dagger }{ n_0 }^{T}
    \!\bigg)
.
\nonumber
\eea
The summation in Eq.~(\ref{momenta:matrix}) runs over $b$, while 
$a$, $i$ and $j$ are held fixed.  

The eigenvalues $\gamma_i$ of the matrix $A^{-1}\big((i/2)B[m_d^2,\cdot] + \Gamma C\big)$ 
determine the damping and oscillatory behavior of the solution.  
Due to the form of the source, the first few momenta dominate the
solution.
If the source has a compact support, we can deduce that outside this compact 
region, $\mu$ is a superposition of damped harmonic oscillations
 with the frequencies ${\Im}(\gamma_i)$ 
and damping rates ${\Re}(\gamma_i)$. 
The amplitude of these oscillations is then suppressed by 
$|\gamma_i|^{-1}$,
such that fast oscillating modes  
give smaller contributions to the current. 

\section{Domination by diagonal parts}
\label{Domination by diagonal parts}

For special choices of the mass matrix, or
in the limit where the oscillations suppress the off-diagonal contribution,
the problem can be reduced to the diagonal case and
the first order contributions to the CP violation, produced by
the oscillations of the off-diagonal terms, are negligibly small.
When viewed in the mass eigenbasis, the problem then reduces
to the diagonal case, such that the first CP-violating contributions
come from the second order semiclassical force in the kinetic equation.
 This approach was originally pursued 
in~\cite{ProkopecSchmidtWeinstock:2003, ProkopecSchmidtWeinstock:2004},
and we summarize its main results in
 section~\ref{Reduction to the diagonal limit}.

We have seen that the first order terms are for large damping 
suppressed as $\Gamma^{-2}$. 
In this section we pose the question how are the second order terms
suppressed in this limit. 

Since $\Gamma$ is large 
we integrate the Taylor expansion of the source using the Green function method and
notice, that the first coefficient gives no contribution (since it is odd in $k_z$)
 and the second term gives
\bea
g^{(2)D}_{0} = \frac{is k_z}{8k_0^2\Gamma^2  \tilde k_0}
  \big\{{m^\dagger}^{\prime\prime} m 
         - m^\dagger m^{\prime\prime} , 
        \partial_{k_z} \hat g^{(0)}_{0}\big\}^{\prime D}
\label{g2:once more}
\,. 
\eea
The term~(\ref{g2:once more})
 is suppressed by $\Gamma^2$ as the contributions in the first order
terms, but in addition by two more orders in the gradient expansion. 
Therefore we can not infer that the second order terms dominate for 
large damping. Rather the region in parameter space
where $\Lambda$ is large leads to 
dominance of the diagonal terms.
However, the second order terms can yield CP violation 
in the trace of the Green function,
while the first order terms are always traceless. 
Therefore the second order terms could 
be more important for generating a baryon asymmetry, 
depending on which contribution is more efficiently
transformed into the BAU. 

\section{Local Contributions to the Currents in the MSSM}
\label{Local Contributions to the Currents in the MSSM}

In this section we give the explicit expressions for the CP-violating 
currents in the MSSM in the local approximation. 
Since in the MSSM we expect the damping $\Gamma$ to be less than ${T_c}/{20}$, 
we are in the regime, where transport is important, 
such that the main intention is to make our approach comparable
 with former publications, in which local sources for diffusion equations
have been derived~\cite{CarenaMorenoQuirosSecoWagner:2000,CarenaQuirosSecoWagner:2002}. 

The most important contribution to the BAU in the MSSM is determined by the 
mass matrix of the chargino-higgsino sector with complex
 $M_2, \mu_c$ and real $H_1,H_2$.
\begin{equation}
  m = \left(\begin{array}{cc}
                    M_2 & gH_2^* \cr
                    gH_1^*  & \mu_c \cr
      \end{array}\right)
\end{equation}
The procedure how to diagonalize $m$ is outlined in Appendix \ref{app:diag}.

 Using this parametrization we can evaluate the CP-violating 
chiral source current $j^{T\slash{Q}}_R$~(\ref{currentsource:local}) 
and $j^{T\slash{Q}}_L$. Since this sources are traceless, the relevant 
quantities are ${\rm Tr}\big(\sigma^3 j^{T\slash{Q}}_R\big)$ and
${\rm Tr}\big(\sigma^3 j^{T\slash{Q}}_R\big)$, where
 $\sigma^3 = {\rm diag}(1,-1)$ in flavour space. 
Under the trace Eq.~(\ref{currentsource:local}) can be reduced to 
the form
\begin{eqnarray}
{\rm Tr}\big(\sigma^3 j_R^{T\slash{Q}}\big)
  &=&   -\frac{is}{8}\!\int_{k_0>0}\frac{d^4k}{(2\pi)^3 }
 \frac{\Lambda}{k_0^2\Gamma^2 \!+\! \frac{1}{4}\Lambda^2} 
           \frac{k_0}{|\tilde k_0|}
 \nn \\
 && \hskip -1 cm \times \,
   \bigg(\!-\!{\rm Tr}\big[\sigma^3\delta(k^2\!-\!|m_d|^2)n\big]
                              {\rm Tr}\big[ (V\sigma^3V^\dagger)^T
                      \big\{
                           4 k_z^2({VV^\dagger}^\prime)^T
                         + \big(V(m\ddp m \!-\! m\dd m^\prime)V^\dagger\big)^T
                       \big\}
                    \big]
\nn\\
&&  \hskip 0.3cm
 +\; {\rm Tr}\big[\delta(k^2\!-\!|m_d|^2)n\big]
     {\rm Tr}\big[ (V\sigma^3V^\dagger)^T
                \big(\sigma^3 V(m\dd m)^\prime V^\dagger\big)^T
             \big]
    \!\bigg)
.
\label{currentsource:local:2}
\end{eqnarray}
 The traces can be easily evaluated by making use of Appendix~A,
\begin{eqnarray}
  {\rm Tr}\big[ (V\sigma^3V^\dagger)^T({VV^\dagger}^\prime)^T\big]
      &=& 4i\frac{1}{\Lambda^2}\Im(M_2\mu_c)
              \big(u_1^\prime u_2 - u_1 u_2^\prime\big)
\nn\\
{\rm Tr}\big[ (V\sigma^3V^\dagger)^T
                         \big(V(m\ddp m \!-\! m\dd m^\prime)V^\dagger\big)^T
                    \big]
      &=& -4i\frac{\bar\Delta}{\Lambda^2}\Im(M_2\mu_c)
                 \big(u_1 u_2\big)^\prime
\nn\\
     {\rm Tr}\big[ (V\sigma^3V^\dagger)^T
                \big(\sigma^3 V(m\dd m)^\prime V^\dagger\big)^T
             \big]
      &=& -4i\frac{1}{\Lambda}\Im(M_2\mu_c)
                 \big(u_1^\prime u_2 - u_1 u_2^\prime\big)
\,,
\end{eqnarray}
where we used $\bar \Lambda = \Lambda$.
The form of the chiral source can be then written as 
(we reinsert the spin superscript)
\begin{eqnarray}
{\rm Tr}\big(\sigma^3 j^{s\slash{Q}}_R\,\big)
      &=& s\frac{\Im(M_2\mu_c)}{T_c^2} \frac{\bar\Delta}{T_c^2}
                   \big(u_1 u_2\big)^\prime \eta_{(0)}^3
                   - s\frac{\Im(M_2\mu_c)}{T_c^2} 
                   \big(u_1^\prime u_2 - u_1 u_2^\prime\big)
                   \big(\eta_{(0)}^0 + 4\eta_{(2)}^3\big)
\label{jsQR}
\\
{\rm Tr}\big(\sigma^3 j^{s\slash{Q}}_L\,\big)
   &=& s\frac{\Im(M_2\mu_c)}{T_c^2} \frac{\Delta}{T_c^2}
                   \big(u_1 u_2\big)^\prime \eta_{(0)}^3
                   + s\frac{\Im(M_2\mu_c)}{T_c^2} 
                   \big(u_1^\prime u_2 - u_1 u_2^\prime\big)
                   \big(\eta_{(0)}^0 + 4\eta_{(2)}^3\big)
\,,
\label{jsQL}
\end{eqnarray}
where $\bar \Delta = |M_2|^2 -|\mu_c|^2 + (u_1^2 - u_2^2)$,
$\Delta = |M_2|^2 -|\mu_c|^2 - (u_1^2 - u_2^2)$, and 
we defined the integrals, 
\bea
\eta_{(n)1/2} &\equiv& 
  T_c^{2-n} \, \int_{k_0>0} \frac{d^4k}{(2\pi)^3} \frac{k_0}{\tilde k_0}
  k_z^n \frac{n(k^\mu,m_{1/2}^2)}
{k_0^2 \Gamma^2 + (\Lambda/2)^2} \delta(k^2 - m_{1/2}^2)
\nn\\
 \eta_{(n)}^0 &\equiv& \frac{1}{2}\big(\eta_{(n)1} + \eta_{(n)2}\big)
\,,\qquad
 \eta_{(n)}^3 \equiv \frac{T_c^2}{2\Lambda}\big(\eta_{(n)1}-\eta_{(n)2}\big) 
\,,
\qquad
\Lambda = m_1^2 - m_2^2
\,.
\label{eta:def}
\eea
The functions $\eta_{(n)}^0$ and $\eta_{(n)}^3$ are dimensionless 
and depend only weakly on $\Lambda$  
in the region where $|k_0|\Gamma \geq \Lambda$, but generate a behavior 
$\propto {T_c^4}/{\Lambda^2}$ in the limit $\Gamma \!\rightarrow\! 0$.
In our former publication~\cite{ProkopecSchmidtWeinstock:2003} we 
neglected off-diagonals, so we were working in the limit 
of large $\Lambda$.

 Several comments are now in order. 
Equations~(\ref{jsQR}--\ref{jsQL}) represent the new CP-violating 
sources which were calculated by solving iteratively the quantum kinetic 
equations for two mixing charginos of the MSSM. 
These sources are absent in the single fermion case, in which 
case the diagonal semiclassical force source dominates. 
Both of the chiral source currents are proportional to spin, and hence 
when summed over spin the sources vanish,
\begin{equation}
 \sum_s j_R^{s\slash{Q}} = 0
\,,\qquad  
 \sum_s j_L^{s\slash{Q}}  = 0
\,.
\label{sources:spin summed}
\end{equation}
Nonvanishing sources are obtained only when a weighted sum over spin 
is performed (for a related discussion of the semiclassical force source 
see Refs.~\cite{ProkopecSchmidtWeinstock:2003,ProkopecSchmidtWeinstock:2004}),
$ \sum_s s j_R^{s\slash{Q}}$ and $ \sum_s s j_L^{s\slash{Q}}$.
To get the source currents for a moving wall, we assume that 
in the plasma frame the current transforms as a Lorentz vector, which 
is reasonable provided diffusion is inefficient. 
 In order to facilitate a comparison with the 
existing work on electroweak baryogenesis sources, it is instructive to 
calculate the vector and axial source currents. 
 From Eqs.~(\ref{jsQR}--\ref{jsQL}) we then easily
 get for the currents in the plasma frame and for a moving wall
\begin{eqnarray}
{\rm Tr}\big(\sigma^3 j_{5\,\mu}^{\slash{Q}}\big)
  &\equiv& {\rm Tr}\sigma^3 \sum_{s} \frac{s}{2}\big({j^{s\slash{Q}}_{R\mu}}
                                   + {j^{s\slash{Q}}_{L\mu}}
                                \big)
  = 2\frac{\Im(M_2\mu_c)}{T_c^2} 
      \frac{|M_2|^2 - |\mu_c|^2}{T_c^2}
             \big[\partial_\mu \big(u_1 u_2\big)\big] 
      \eta_{(0)}^3
\label{jsQ0}
\\
{\rm Tr}\big(\sigma^3 j_\mu^{\slash{Q}}\big)
   &\equiv& {\rm Tr}\sigma^3 \sum_{s}\frac{s}{2}\big(j_{R\mu}^{s\slash{Q}}
                             - j_{L\mu}^{s\slash{Q}}
                          \big)
         = 2\frac{\Im(M_2\mu_c)}{T_c^2} 
            \frac{u_1^2-u_2^2}{T_c^2}
       \big[\partial_\mu(u_1 u_2)\big] \eta_{(0)}^3
\nonumber\\
  &&   \hskip 4.62cm
     -\;   2\frac{\Im(M_2\mu_c)}{T_c^2} 
                   \big( u_2\partial_\mu u_1 - u_1\partial_\mu u_2\big)
                   \big(\eta_{(0)}^0 + 4\eta_{(2)}^3\big)
\,,\qquad
\label{jsQ5}
\end{eqnarray}
such that the sources in the local approximation neatly split
 into the plus and minus 
contributions, $\propto \partial_\mu(u_1 u_2)$
and $\propto  u_2\partial_\mu u_1- u_1\partial_\mu u_2$, respectively.
The axial current is sourced by the plus contribution only
(just like in the case of the second order semiclassical force),
while the vector current is sourced by both plus and minus contributions. 
 In the non-local case, both plus and minus terms contribute to
$j^{\slash{Q}}_\mu$ and $j^{\slash{Q}}_{5\mu}$.
These results have a similar structure to the sources found in 
Refs.~\cite{CarenaMorenoQuirosSecoWagner:2000} 
and~\cite{CarenaQuirosSecoWagner:2002}. The differ however, 
when a detailed quantitative comparison is made.

Finally, we quote the second order diagonal source calculated 
in the local approximation~(\ref{g2:once more}):
\begin{equation}
{\rm Tr}\big(\mathbbm{1} \, j_{5\,\mu}^{(2)}\big)
   \simeq 2\frac{\Im(M_2\mu_c)}{T_c^4}
      \partial_\mu\Big( u_1^{\prime\prime}u_2+u_1u_2^{\prime\prime}
      \Big) \zeta_{(0)}^3
\label{j02:CP}
\,
\end{equation}
with the definitions
\bea
\zeta_{(n)1/2} &\equiv& 
  T_c^{2-n} \, \int_{k_0>0} \frac{d^4k}{(2\pi)^3} \frac{k_0}{\tilde k_0}
  k_z^n \frac{n(k^\mu,m_{1/2}^2)}{k_0^2 \Gamma^2} 
\delta(k^2 - m_{1/2}^2)
\,,\quad
\zeta_{(n)}^3 \equiv \frac{T_c^2}{2\Lambda}\big(\zeta_{(n)1}-\zeta_{(n)2}\big) 
.
\qquad
\label{zeta:def}
\eea
Note that, in contrast to $\eta$, $\zeta$ is not suppressed for large
$\Lambda$, such that the second order terms dominate in the local regime 
for large values of $\Lambda$.

\section{Numerical Results of Transport in the MSSM}
\label{Numerical Results of Transport in the MSSM}

In this section we will present numerical results of the fluid ansatz.
The Higgs {\it vevs} and the $\beta$ angle are parametrized by $H_1(z) = H(z) \sin(\beta(z))$,
$H_2(z) = H(z) \cos(\beta(z))$ and  
\bea
H(z) = \frac{1}{2} v(T) 
  \lp 1- \tanh \lp \alpha \lp 1- \frac{2z}{L_w}\rp\rp \rp, 
\label{H}
\\
\beta(z) = \beta_\infty - \frac{1}{2} \Delta\beta 
\lp 1+ \tanh \lp \alpha \lp 1- \frac{2z}{L_w}\rp\rp \rp. 
\label{beta}
\eea
If not stated differently, the parameters used
in the plots are $T_c=95$ GeV, $v(T)=175$~GeV, $\alpha=\frac{3}{2},\,
\tan(\beta_\infty)=10,\, T_c \, L_w=20,\, \Gamma = \alpha_W T_c$,  
$M_2=200$ GeV, $\mu_c=250$ GeV,
 and the complex phase is chosen maximally $\Im(M_2 \mu_c)=|M_2 \mu_c|$.
The value of $\Delta\beta = 0.0108$ is deduced from \cite{Moreno:1998bq} by 
using the value $m_A=200$ GeV. 
In figure~\ref{fig:H+beta} we show how the wall~(\ref{H}--\ref{beta})
 looks for our choice of parameters. 
\begin{figure}[tbp]
\centerline{
\epsfig{figure=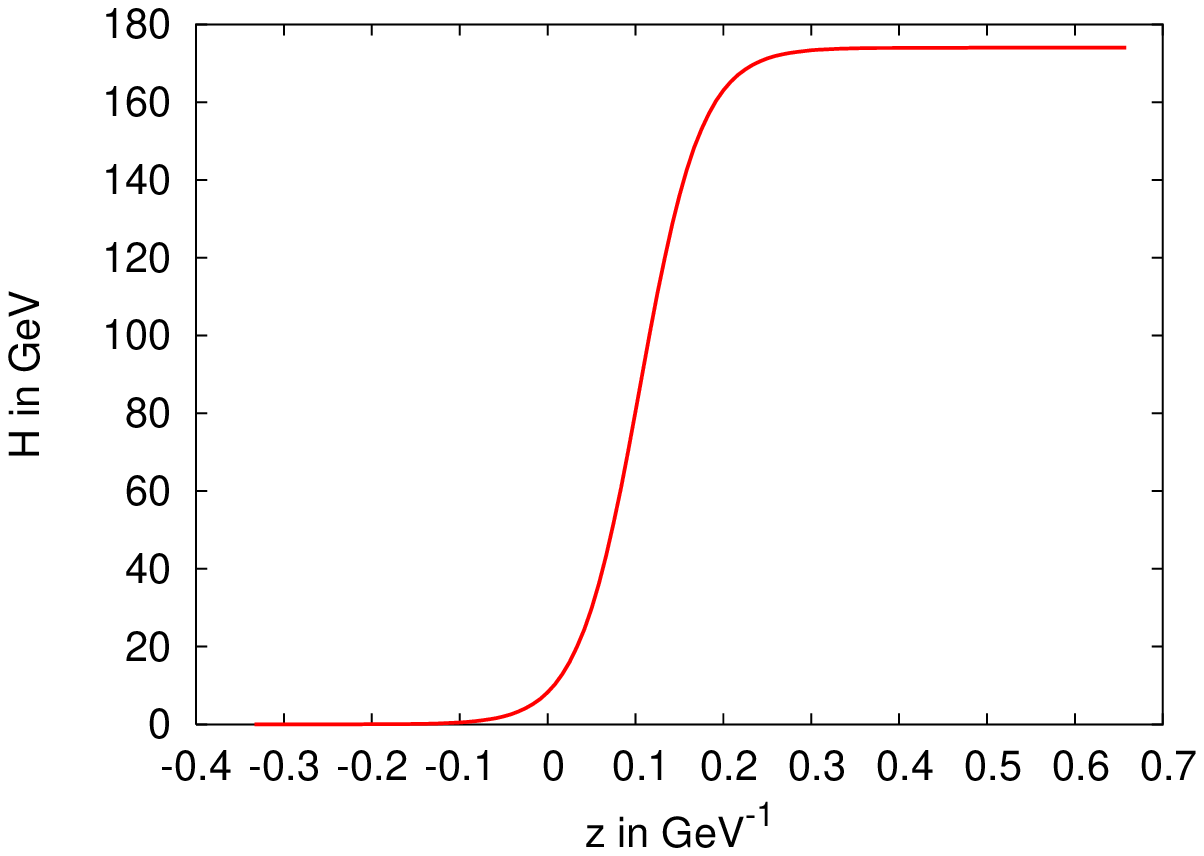, height=3.in,width=3.3in}
\epsfig{figure=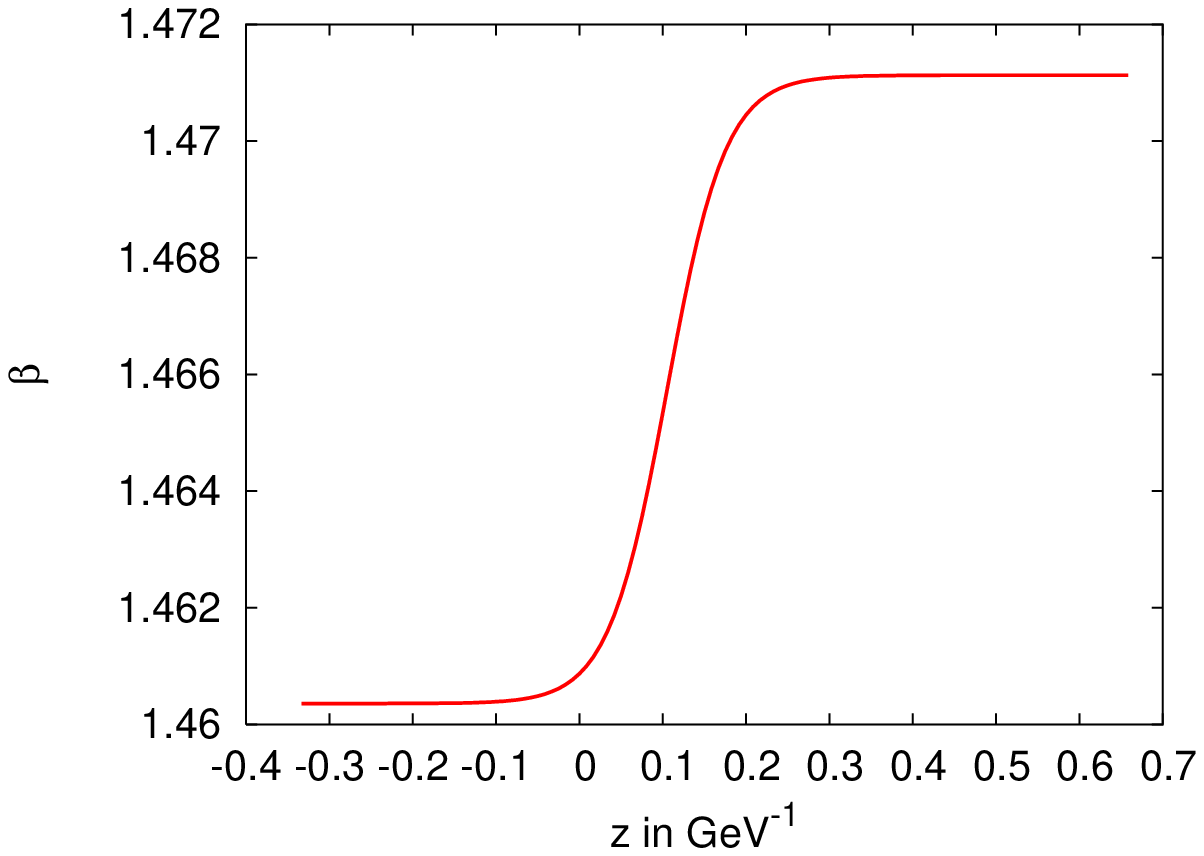, height=3.in,width=3.3in}
}
\caption{\small
{\it Bubble wall:}
 The higgs {\it vev} profile~(\ref{H}) and $\beta$~(\ref{beta}) as 
a function of $z$. 
}
\lbfig{fig:H+beta}
\end{figure}

The wall velocity is taken to be $v_w=0.05$ and
 the plots are evaluated for the first six momenta. 
The currents $j^0$, $j^0_5$ and the second order term ${j_5}^{(2)}$
are evaluated in the plasma frame, thus the expressions
are linear in $v_w$. 

First we will display the dependence on the number of momenta that are
used in our fluid {\it Ansatz}.
 Fig. \ref{Fig:N-dependence} shows that the convergence is 
already very good for $N=6$, such that it is sufficient to use only the first six momenta.
\begin{figure}[htbp]
\begin{center}
\epsfig{file=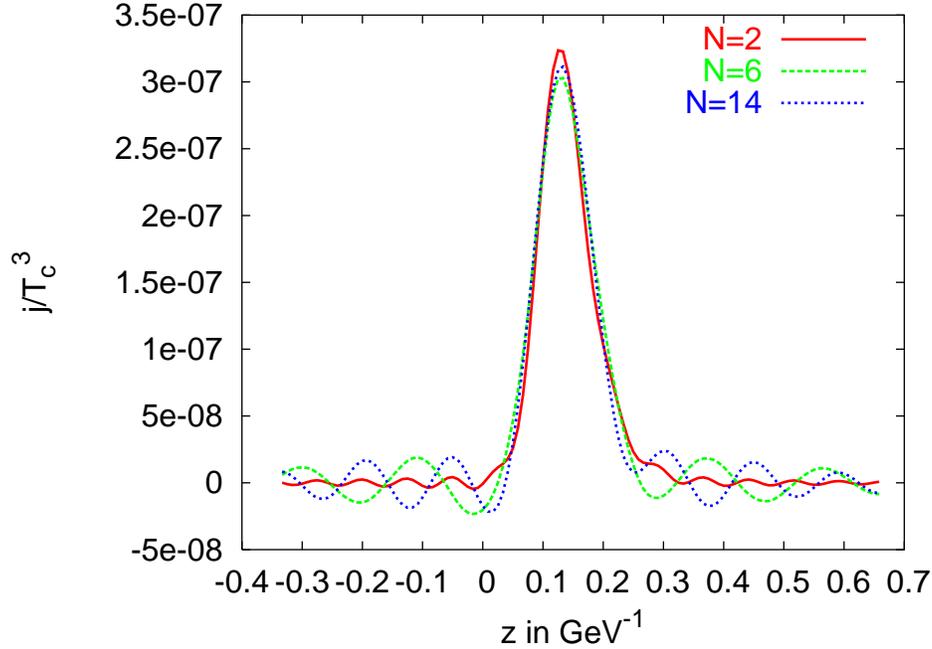, height=3.5 in,width=5 in}
\end{center}
\vskip -0.2in
\lbfig{Fig:N-dependence}
\caption[Fig:N-dependence]{%
\small
Dependence of the current for a typical source on the number of momenta $N$ used. 
}
\end{figure}
Since the higher momenta have a slightly smaller exponential suppression factor
but much smaller sources, these contributions become important in the region 
far away from the source, where the oscillations take place. 
Mostly the phases of the oscillations are influenced, such that quantitative
statements barely change for $N>6$.

The plots Fig. \ref{Fig:muc_205} to Fig. \ref{Fig:muc_450} show 
all three currents for the values $\mu_c=\{205,220,250,450\}$ GeV
and $M_2=200$ GeV. For small $\Lambda$, respectively $M_2 \simeq \mu_c$, 
the solution is oscillating and has rather large amplitudes.  
These oscillations are, as expected, 
suppressed for larger values of $\Lambda$ and a local 
contribution remains. For $\Lambda \sim 20\, T_c^2 \sim T_c^3L_w$ the 
second order contribution, which shows a weaker dependence on $\Lambda$, 
begins to dominate. When $\Lambda$ is large, 
the first order currents are suppressed due to efficient decoherence.
In the BAU, the second order terms start to dominate earlier
since, for the first order terms, the oscillations and inefficient transport 
prevent in part an efficient source conversion to baryon asymmetry, 
while the second order terms are transported more efficiently
and without oscillations, such that that give a truly non-local contribution.

The term resulting from the combination 
$u_1 \partial_\mu u_2 - u_2 \partial_\mu u_1$ is 
suppressed due to the fact
that $\Delta\beta$ is small. In addition, the terms that include  
$\eta^3$ are smaller than the terms including $\eta^0$. 

\begin{figure}[htbp]
\epsfig{file=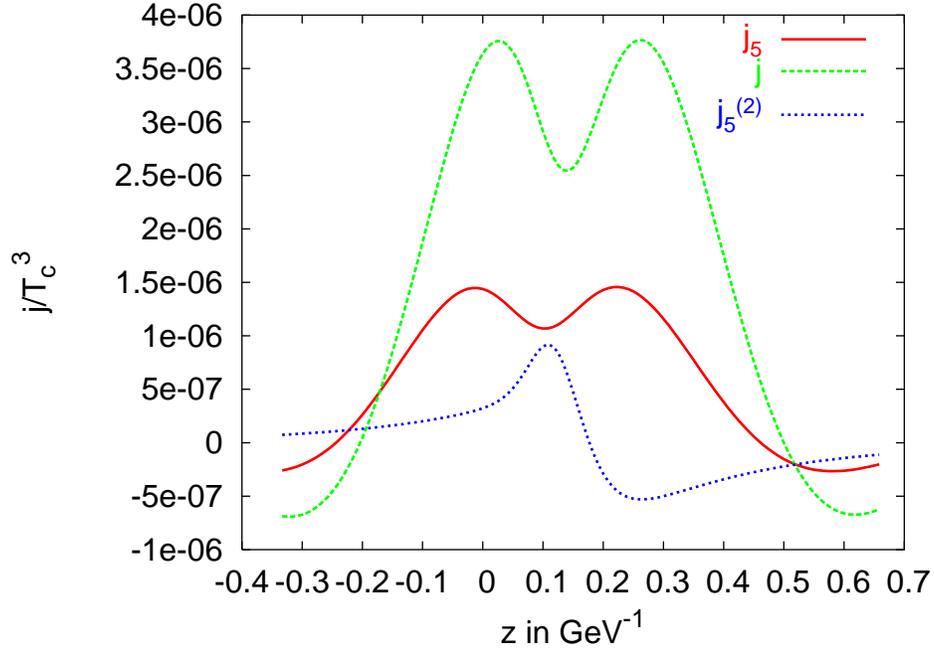, height=3.5 in,width=5 in}
\vskip -0.2in
\lbfig{Fig:muc_205}
\caption[Fig:muc_205]{%
\small
The CP-violating currents of first order 
$j^0$, $j^0_5$ and of second order ${j_5^0}^{(2)}$.
$M_2=200$ GeV. 
$\mu_c=205$ GeV, $\Lambda/T_c^2 \in [0.22, 4.0]$. 
}
\end{figure}
\begin{figure}[htbp]
\epsfig{file=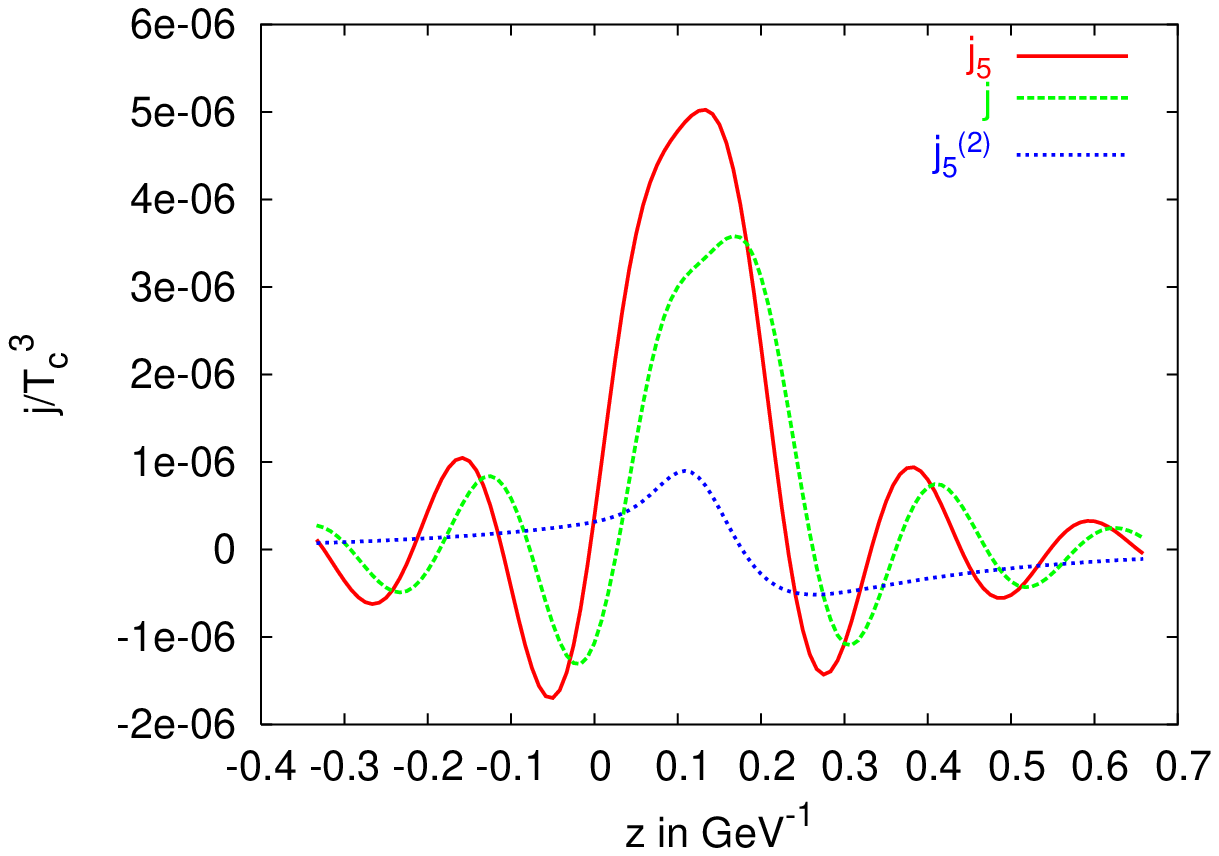, height=3.5 in,width=5 in}
\vskip -0.2in
\lbfig{Fig:muc_240}
\caption[Fig:muc_240]{%
\small
The CP-violating currents of first order 
$j^0$, $j^0_5$ and of second order ${j_5^0}^{(2)}$.
$M_2=200$ GeV. 
$\mu_c=220$ GeV, $\Lambda/T_c^2 \in [0.93, 4.2]$. 
}
\end{figure}
\begin{figure}[htbp]
\epsfig{file=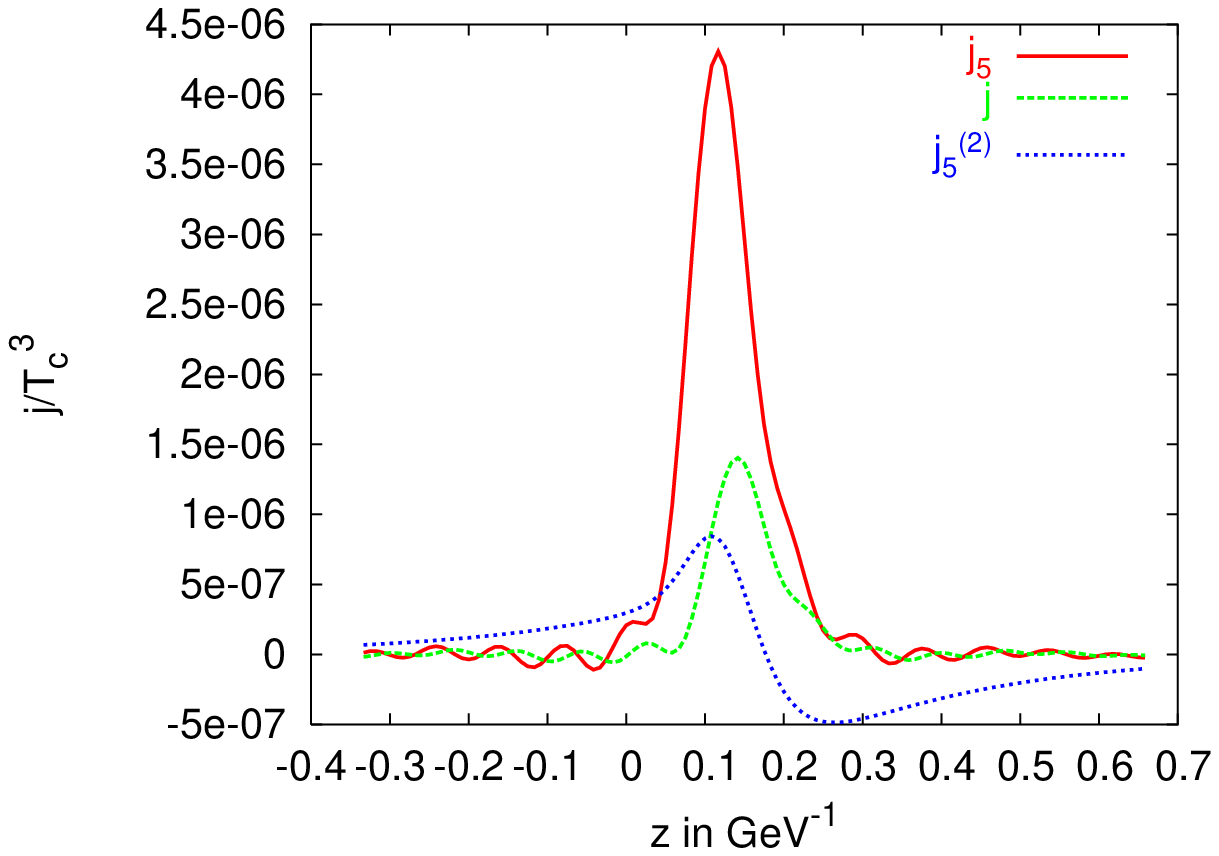, height=3.5 in,width=5 in}
\vskip -0.2in
\lbfig{Fig:muc_300}
\caption[Fig:muc_300]{%
\small
The CP-violating currents of first order 
$j^0$, $j^0_5$ and of second order ${j_5^0}^{(2)}$.
$M_2=200$ GeV. 
$\mu_c=250$ GeV, $\Lambda/T_c^2 \in [2.4, 5.1]$. 
}
\end{figure}
\begin{figure}[htbp]
\epsfig{file=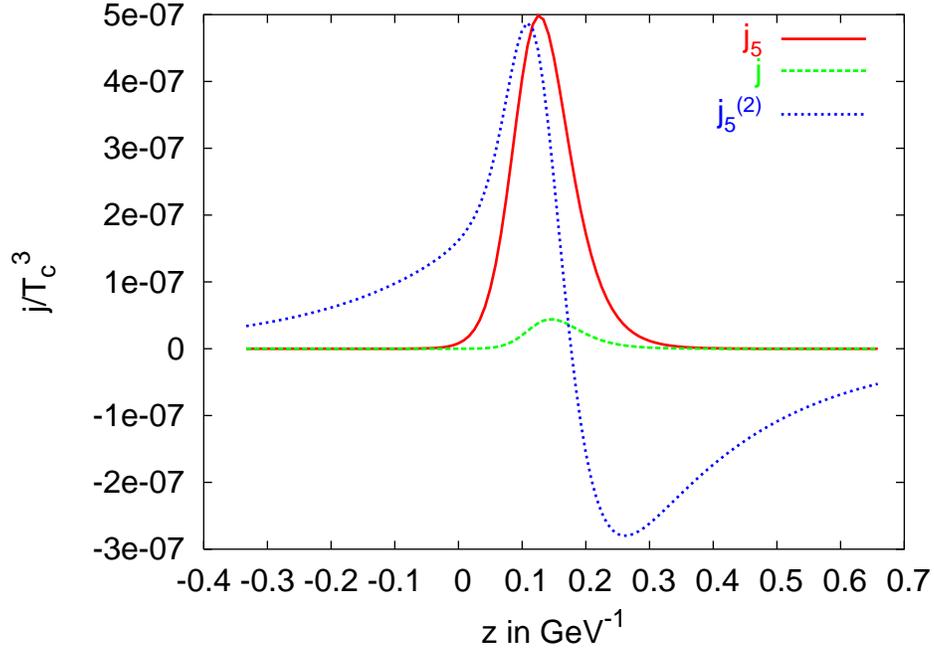, height=3.5 in,width=5 in}
\vskip -0.2in
\lbfig{Fig:muc_450}
\caption[Fig:muc_450]{%
\small
The CP-violating currents of first order 
$j^0$, $j^0_5$ and of second order ${j_5^0}^{(2)}$.
$M_2=200$ GeV. 
$\mu_c=450$ GeV, $\Lambda/T_c^2 \in [18.0, 19.2]$. 
}
\end{figure}
The axial vector current $j^0_5$, that is normally the largest 
contribution to the source,
is suppressed for small $\Lambda$ due to the factor
 $|M_2|^2-|\mu_c|^2$ in Eq.~(\ref{jsQ0}) and in this 
region the vector current $j^0$ becomes the most important one.

\section{Discussion and Outlook}

In this article we have presented a method to solve Schwinger-Dyson equations
for CP-violating densities
for mixing fermions in a space time dependent background.
The transition to the chiral basis was important for the partial decoupling
of the different coefficients in spinor space. 
The terms that can appear in the kinetic equations 
written in chiral basis all have the same 
transformation properties under flavor basis transformations. 
As an intermediate result, we have obtained
the formally exact equations~(\ref{KE:gL:exact}),
in which only two of the 16 coefficients in spinor space remain coupled to
each other.

Next, we have found that the off-diagonal densities exhibit oscillations
at lowest order in gradient expansion.
Even though they vanish without space-time dependent background, 
these terms ought to be treated by taking into account oscillations 
as soon as they are sourced.

In section~\ref{CP-violating sources} we advance a novel definition of 
CP violation in kinetic equations with mixing flavors. According 
to our definition, when kinetic equations with mixing fermions are
truncated at first order in gradients, only the inclusion of flavor 
oscillations (formally expressed through the commutator term) gives raise
to nonvanishing CP violation in the diagonal entries of the flavor basis.
The same is of course true for the chargino and neutralino sectors
in supersymmetric extensions of the Standard Model. This is the main
difference between our approach and the approach advocated in  
literature~\cite{CarenaQuirosRiottoViljaWagner:1997,CarenaMorenoQuirosSecoWagner:2000,CarenaQuirosSecoWagner:2002}.
Without taking the flavor oscillations into account,
the CP-violating densities would stay in the
off-diagonal entries even after rotation into the flavor basis.

Our approach to second order diagonal sources
(semiclassical force) differs from the treatment  
advocated in Refs.~\cite{ProkopecSchmidtWeinstock:2003,
ProkopecSchmidtWeinstock:2004,ClineJoyceKainulainen:2000+2001}. 
In order to arrive at a local analytic estimate in flavor basis, 
we have considered the limit of large damping, $\Gamma L\gg 1$.
The CP-violating axial current~(\ref{j02:CP}) is in this case 
proportional to the trace in flavor space.
The source from semiclassical force
 in~\cite{ProkopecSchmidtWeinstock:2003,
ProkopecSchmidtWeinstock:2004,ClineJoyceKainulainen:2000+2001}
is calculated in the mass eigenbasis in the limit when $\Gamma L\ll 1$,
and it was found to be proportional to the difference of flavor
axial densities,
${\rm Tr}(\sigma^3 j^3_{5d})$. Moreover, in our numerical treatment
we calculate the source by taking account of both flavor mixing
and transport, while the same source has been treated
in the literature in the diagonal approximation in the mass eigenbasis.

Apart from the {\it plus} contribution,
$\propto u_1 \partial_\mu u_2 + u_2\partial_\mu u_1
    \equiv \partial_\mu(u_1u_2)$,
we also found the {\it minus} contribution, 
$u_1 \partial_\mu u_2 - u_2 \partial_\mu u_1$.
The plus contribution is sourced by both the first order off-diagonals
and by the second order diagonals. 
The minus term plays an important role in 
the approach advocated in
Refs.~\cite{CarenaMorenoQuirosSecoWagner:2000,CarenaQuirosSecoWagner:2002},
especially near the degeneracy (small $\Lambda$), where it exhibits
a resonant enhancement. When compared with our results,
in the region of near degeneracy we find a weak enhancement in all
contributions to the CP-violating vector current, such that the minus 
contribution remains subdominant.

By performing a numerical study of fluid equations, we have analysed
the CP-violating vector and axial vector currents for a slowly moving wall, 
and found that, in the nonlocal regime, in which the
currents are only weakly damped $\Gamma L_w \le 1$, the first order terms 
provide a dominant contribution to the CP-violating currents if
$\Lambda = m_i^2 - m_j^2$ ($m_i^2$ ($i = 1,..,N$) denote the mass eigenvalues)
is smaller than about $20 \, T_c^2$.
Whether this statement remains true with respect to the baryon asymmetry 
remains unclear, since the oscillations, poor transport
and the tracelessness of the first order terms 
could prevent an efficient production of BAU. 

 A more comprehensive comparison with the work of
Refs.~\cite{CarenaMorenoQuirosSecoWagner:2000,CarenaQuirosSecoWagner:2002} 
and~\cite{ProkopecSchmidtWeinstock:2003,
ProkopecSchmidtWeinstock:2004,ClineJoyceKainulainen:2000+2001},
and the explicit calculation of the BAU
is postponed to a forthcoming publication. 

\section*{Acknowledgments}

We would like to thank Steffen Weinstock and Marco Seco for numerous useful
discussions.

\appendix 

\section{Diagonalization of the Chargino-Higgsino mass matrix \label{app:diag}}

The chargino-higgsino mass matrix is given by
\begin{equation}
  m = \left(\begin{array}{cc}
                    M_2 & gH_2^* \cr
                    gH_1^*  & \mu_c \cr
      \end{array}\right)
\end{equation}
The mass matrix $m$ is diagonalized by the biunitary transformation
\begin{equation}
  m_d = UmV^\dagger
\,,
\end{equation}
with 
\begin{eqnarray}
  U &=& \left(\frac{2}{\Lambda(\Lambda+\Delta)}\right)^\frac 12
      \left(\begin{array}{cc}
                    \frac 12 (\Lambda + \Delta) & a \cr
                    -a^*  & \frac 12 (\Lambda + \Delta) \cr
            \end{array}
      \right)
\label{U}
\\
 a &=& g(M_2H_1 + \mu_c^* H_2^*) 
\,,\qquad
 \Delta = |M_2|^2 - |\mu_c|^2 - (u_1^2 - u_2^2)
\,,\qquad
 \Lambda = (\Delta ^2 + 4|a|^2)^\frac 12
\nonumber
\end{eqnarray}
and
\begin{eqnarray}
  V &=& \left(\frac{2}{\bar\Lambda(\bar\Lambda+\bar\Delta)}\right)^\frac 12
      \left(\begin{array}{cc}
                    \frac 12 (\bar \Lambda + \bar\Delta) & \bar a \cr
                    -\bar a^*  & \frac 12 (\bar\Lambda + \bar\Delta) \cr
            \end{array}
      \right)
\label{V}
\\
 \bar a &=& g(M_2^*H_2^* + \mu_c H_1) 
\,,\qquad
 \bar \Delta = |M_2|^2 - |\mu_c|^2 + (u_1^2 - u_2^2)
\,,\qquad
 \bar\Lambda = (\bar\Delta ^2 + 4|\bar a|^2)^\frac 12  =  \Lambda
\nonumber
\end{eqnarray}
where we defined $u_{1,2} = |g H_{1,2}|$.
Note that $\bar a$ and $\bar\Delta$ can be obtained from $a$ and $\Delta$
by the replacements, $M_2\leftrightarrow M_2^*$, 
$\mu_c\leftrightarrow \mu_c^*$ and $H_1\leftrightarrow H_2^\dagger$,
such that $\bar\Lambda = \Lambda$, as indicated in~(\ref{V}).
The mass eigenvalues-squared are given by 
\begin{equation}
  {m_d}_{1/2}^2 = \frac 12 \big(|M_2|^2 + |\mu_c|^2 + (u_1^2 + u_2^2)\big)
                \pm \frac{\Lambda}{2}
\label{md:eigenvalues}
\end{equation}
and can be calculated quite simply by noting 
that $U mm^\dagger U^\dagger = m_d^2 = V m^\dagger m V^\dagger$.

\end{document}